\documentclass[pre,aps,twocolumn,superscriptaddress,showpacs,floatfix]{revtex4-2}
\usepackage{graphicx}
\usepackage{amssymb}
\usepackage{natbib}
\usepackage{dcolumn}
\usepackage{bm}
\usepackage{color}
\usepackage[colorlinks=true, linkcolor=blue, citecolor=blue]{hyperref}
\usepackage{subfigure}
\usepackage{graphicx,psfrag,xspace}
\usepackage{color}
\usepackage{amssymb,amsfonts,amsmath}
\usepackage{setspace}
\usepackage[normalem]{ulem}  

\newcommand{\eq}[1]{(\ref{#1})}

\newcommand{\gdot}{\dot{\gamma}}
\newcommand{\sigmaY}{{\sigma_{\tt Y}}}

\newcommand{\br}{{\bm r}}

\begin{document}

\author{E. E. Ferrero} 
\affiliation{Instituto de Nanociencia y Nanotecnolog\'{\i}a, CNEA--CONICET, 
Centro At\'omico Bariloche, (R8402AGP) San Carlos de Bariloche, R\'{\i}o Negro, Argentina.}

\author{A. B. Kolton}
\affiliation{Centro At\'omico Bariloche, Instituto Balseiro, 
Comisi\'on Nacional de Energ\'ia At\'omica, CNEA, CONICET, UNCUYO,
Av. E. Bustillo 9500 R8402AGP S. C. de Bariloche, R\'io Negro, Argentina}

\author{E. A. Jagla}
\affiliation{Centro At\'omico Bariloche, Instituto Balseiro, 
Comisi\'on Nacional de Energ\'ia At\'omica, CNEA, CONICET, UNCUYO,
Av. E. Bustillo 9500 R8402AGP S. C. de Bariloche, R\'io Negro, Argentina}

\title{The yielding of amorphous solids at finite temperatures}

\begin{abstract}
We analyze the effect of temperature on the yielding transition of amorphous solids 
using different coarse-grained model approaches.
On one hand we use an elasto-plastic model, with temperature introduced in the form 
of an Arrhenius activation law over energy barriers.
On the other hand, we implement a Hamiltonian model with a relaxational dynamics,  
where temperature is introduced in the form of a Langevin stochastic force.
In both cases, temperature transforms the sharp transition of the athermal case in a 
smooth crossover.
We show that this thermally smoothed transition follows a simple scaling form that 
can be fully explained using a one-particle system driven in a potential under the 
combined action of a mechanical and a thermal noise, namely the stochastically-driven 
Prandtl-Tomlinson model. 
Our work harmonizes the results of simple models for amorphous solids
with the phenomenological $\sim T^{2/3}$ law proposed by Johnson and Samwer 
[Phys. Rev. Lett. 95, 195501 (2005)] in the framework of experimental 
metallic glasses yield observations, and extend it to a generic case.
Finally, our results strengthen the interpretation of the yielding transition
as an effective mean-field phenomenon.
\end{abstract}

\maketitle

\section{Introduction}


Amorphous materials are neither perfect solids nor simple liquids. 
Foams and emulsions, colloidal glasses, oxide and metallic glasses,
glassy polymers and some granular media preserve at rest a solid 
structure, but will flow if a sufficiently large load is applied to them.
Accordingly, in the rheology of complex fluids~\cite{BonnRMP2017}
they are often referred to as ``yield-stress materials''.
The transition between the solid-like elastic response and 
the irreversible plastic deformation is known as the yielding
transition~\cite{NicolasRMP2018}. 
Statistical physicists have regarded it as a dynamical 
out-of-equilibrium phase transition, similar to the 
depinning transition of elastic manifolds in random 
media~\cite{wiese2021theory}, and under the light of 
equilibrium phase transitions theory. 
Notably, the bulk of recent theoretical work on the 
yielding transition of amorphous solids has been devoted 
to the case in which the effect of temperature is disregarded.
Both elastoplastic models and molecular dynamic simulations have
focused on describing and understanding the \emph{athermal} 
deformation and related critical phenomena~\cite{LernerPRE2009,LinPNAS2014,NicolasRMP2018}.
Alike the effect of a small external magnetic field in the 
ferromagnetic-paramagnetic transition of a magnet (say, Ising model), 
when a finite temperature is taken into account in the deformation of
amorphpus solids  it is expected to round up the yielding transition, 
as it does in depinning~\cite{AgoritsasPB2012,FerreCRP2013}.

%
When the elementary constituents of the material are large 
enough ($\gtrsim 1\mu m$) to neglect Brownian motion effects,
an athermal approach is well justified and can be even 
quantitatively predictive (e.g., in dense granular suspensions, 
dry granular packings, foams, and emulsions)
\footnote{Notice that temperature can manifest itself in dependencies 
of intrinsic properties~\cite{NicolasRMP2018} of the material, 
like for example `average bubble size' in a foam, even when there 
is no relevant `agitation' or thermal activation.
When we say `athermal' here we mean no relevant thermal motion.}.
Yet, thermal fluctuations may play a role in materials 
with small enough ($\lesssim 1\mu m$) elementary constituents,
e.g., colloidal and polymeric glasses, 
colloidal gels, silicate and metallic glasses.
For those materials, thermally activated events cannot be immediately 
disregarded. 
It happens typically though, that driven systems respond on 
much shorter times than quiescent aging systems; and then, 
some thermal materials may be treated as athermal for all 
practical purposes when considering mechanical deformation.
Nevertheless, the most interesting physical behavior emerges 
when the thermal agitation and driving time scales compete, 
either because temperature is high enough or because the 
driving is slow. 
The yielding transition, the limit of vanishing strain rate itself,
is of course within this scope.

%
In a famous paper~\cite{JohnsonPRL2005} Johnson and Samwer (J\&S)
analysed the behavior of a broad range of metallic glasses,
finding a universal 
temperature correction to the compressive yield strength
scaling as $\sim T^{\frac23}$:
\begin{equation}
    \tau_{cT} = \tau_{c0} \left(1 - [A T \ln(\omega_0 / C\gdot) ]^\frac23 \right) 
    \label{eq:JandS}
\end{equation}
where $\tau_{cT}$ is the compression stress at yielding
at temperature $T$, and $\tau_{c0}$ the corresponding value at $T=0$.
This law was derived by estimating the transition rate over typical energy 
barriers in a Frenkel-like construction for the elastic energy of shear transformation zones~\cite{JohnsonPRL2005}, 
using an attempt frequency $\omega_0$, and a typical height of barriers that decreases 
as the applied stress is increased, vanishing as a $\frac32$ power of the stress 
remaining to reach instability
\footnote{Note that $\tau_{cT}$ is the applied stress at which a minimum threshold 
strain rate deformation $\dot\gamma$ is experimentally detected, and may or may not 
correspond to a steady-state stress producing a steady-state strain rate.
In general, there is a `stress overshoot'~\cite{benzi2021stress} in the deformation
of soft glassy materials, which depends on strain rate, aging and sample preparation.
Therefore, the (dynamical) yield stress, i.e., the steady-state stress in a quasistatic
deformation, is different from the stress at the onset of yielding.
Yet, in most (if not all) of the works on metallic glasses
cited in Ref.~\cite{JohnsonPRL2005}, the data correspond to `poorly 
annealed' systems. 
The stress at the onset of yielding in such systems, defined at the deviation from the 
elastic regime, is itself already very similar to the stress value expected in an
extrapolated steady state, as no stress overshoot is observed in the data. 
Thus, we take the freedom to interpret the finding of~\cite{JohnsonPRL2005}
in the steady-state context of our work.
}.

The J\&S $\sim T^{\frac23}$ law was recovered by MD simulations of 
2D Lennard-Jones glasses in Ref.~\cite{ChattorajPRL2010}.
This time, not just the stress at yield but the full flow/curve
was tested for thermal effects and compared with the athermal case.
In that numerical work, the following law was proposed and shown to hold 
for the steady-state stress $\sigma$ as a function of $T$ at a stationary 
value of $\gdot$
\begin{equation}
    \sigma(\dot{\gamma},T) = \sigma_c + A_1\sqrt{\dot{\gamma}} - 
    A_2 T^\frac23 [\ln(A_3 T^\frac56 / \dot{\gamma}) ]^\frac23 .
\label{chattoraj}
\end{equation}
Finally, a refinement of the theoretical derivation for the 
$\sim T^{\frac23}$ law was proposed in \cite{DasguptaPRB2013},
basically following the same principle of Arrhenius-like 
activation of Eshelby events, with barriers $U_b \sim (1-\sigma/\sigma_c)^{3/2}$.
In the present work we will interpret Johnson and Samwer's $\sim T^{\frac23}$ 
law as a particular case of our derived scaling laws for the thermal rounding 
of yielding, which are not restricted to a sole kind of energy barriers.

Some endeavors in understanding thermal effects in the deformation of amorphous 
solids proceeded along the path of analyzing the elementary plastic 
events and their temperature dependence~\cite{Schall2007,HentschelPRL2010,CaoPRE2013}.
But the vast majority of literature devoted to the statistical aspects of
the yielding transition (e.g.~\cite{LernerPRE2009, Karmakar2010, KarmakarPRE2010b,LinEPL2014, LinPNAS2014, budrikis2015universality, tyukodi2015depinning, liu2015driving,FerreroSM2019,FerreroPRL2019, FernandezAguirrePRE2018}) has largely ignored the thermal case.
Only very recently a preprint by Popovic and co-workers appeared discussing 
the thermal rounding of the yielding transition~\cite{popovic2020thermally}.
They show that indeed a scaling law for the thermal rounding holds 
in numerical simulations and prove it analytically for 
the H\`ebraud-Lequeux model. 
Interestingly, the thermal rounding scaling with roots in mean-field theory
of charge-density waves depinning~\cite{FisherPRL1983,MiddletonPRB1992}
works very well in spatially distributed systems for the description of yielding, 
essentially with no corrections. 
While this is somehow good news from the phenomenological point of view
(since it simplifies the physical laws to be considered in more applied fields),
it contrasts with the yielding theories claiming non-trivial correlations and 
corrections to scaling in finite dimensions~\cite{LinPRX2016,LinPRE2018}.

In this work, we will discuss along these lines, 
hoping to bring some light to the latter issue.
First, we recall the behavior of thermal rounding in a well studied 
depinning case~\cite{kolton2020thermally},
to show that indeed in short-ranged interaction systems corrections to
scaling are expected. 
After that, we present the possible scenario for thermal rounding of yielding 
obtained by a generalization of the arguments used in depinning. 
Then, we present results of numerical simulations on two different 
coarse-graining frameworks that have been proposed to study the yielding transition.
One is the familiar case of elasto-plastic models that have been used to describe 
yielding for quite a long time already~\cite{NicolasRMP2018}.
We introduce temperature in this models as an Arrhenius activation probability 
over finite energy barriers.
The second framework we consider is a Hamiltonian model in which many mean-field like 
characteristics of yielding have been discussed in recent 
years~\cite{FernandezAguirrePRE2018, FerreroPRL2019}
Finally, we analyze the results we obtain in both of these extended systems 
to the light of a one-particle `mean-field'-like model, the Prandtl-Tomlinson 
model of friction~\cite{Popov2014} with stochastic driving~\cite{jagla2018prandtl}.
The thermal rounding behavior of this model also 
displays~\footnote{See~\cite{MuserPRB2011} for the case without stochastic driving, 
particularly  Eqs. A20 and A39, and also ~\cite{MuserUrbakhRobbins}.}
a $\sim T^{\frac23}$ phenomenological law analogous 
to the one proposed by Johnson and Samwer~\cite{JohnsonPRL2005}. 


\subsection{The thermal rounding scaling}

In standard critical phenomena a symmetry breaking external field transforms 
a sharp transition into a crossover.
For the paradigmatic ferromagnetic-paramagnetic equilibrium 
phase transition the magnetization ($m$) as a function of 
temperature ($T$) and magnetic field ($h$) satisfies (sufficiently 
close to the critical point $T=T_c$, $h=0$) the following scaling relation
\begin{equation}
m (T,h)=  h^a F((T-T_c)/h^{1/b})
\end{equation}
with $F$ a universal scaling function.
The critical exponent $a$ quantifies the effect of magnetic field right at $T_c$. 
In the limit of $h\to 0$ this expression must become field-independent, 
and then, it reduces to the critical form $m\sim (T-T_c)^\beta$ with $\beta=ab$. 
For the depinning transition of elastic manifolds, a thermal rounding scaling expression
was proposed long time ago by Fisher~\cite{Fisher1985,FisherPR1998} (and numerically 
tested by Middleton~\cite{MiddletonPRB1992}) based on the analogy with equilibrium phase transitions.
With the velocity $v$ as the order parameter, the force $f$ as the control parameter 
and the temperature $T$ as a ``symmetry-breaking field'' destroying the pinned phase,
it has the form
\begin{equation}
v(f,T)=  T^\psi G((f-f_c)/T^{1/\alpha}). 
\label{eq:fisher_thermalrounding}
\end{equation}
As for standard phase transitions, a new exponent $\psi>0$ is introduced, 
describing the smearing effect, $v\sim T^{\psi}$ at $f = f_c$. 
The form of the scaling function $G$ is such that for $T\to 0$ we re-obtain the 
expected critical behavior $v\sim (f-f_c)^\beta$ with $\beta=\alpha\psi$ at $T=0$. 
The driving force in the depinning transition thus plays the role of the temperature 
in the magnetic system, and  temperature the role of  external magnetic field. 
Eq.~(\ref{eq:fisher_thermalrounding}) can be shown to rigorously apply in the
fully-connected mean-field problem~\cite{Fisher1985}, or equivalently, in the 
problem of a single particle driven on a disordered potential~\cite{ambegaokar1969,bishop1978,purrello2017}.
Yet, in the more standard situation for the  depinning problem, namely 
short-range elasticity of the manifold in finite dimensions, 
the precise assessment of the thermal rounding has proved to be non-mean-field
and tricky~\cite{ChauvePRB2000}.
For instance, the numerically determined value of the $\psi$ exponent varies widely 
among different models~\cite{Nowak_1998,Roters1999, BustingorryEPL2007, bustingorryPhysB2009, BustingorryPRE2012, Xi2015, purrello2017}.
Furthermore, the scaling form~\cite{Nattermann2001} and its universality has 
been questioned~\cite{MiddletonPRB1992}. 
Even if $\psi$ was a universal exponent, it is not clear whether it is an 
independent exponent or it is related to other depinning exponents. 
More recently, general arguments have suggested that Eq.~(\ref{eq:fisher_thermalrounding}) may be not generic, 
but rather a special case~\cite{kolton2020thermally}, and recent works show that 
elastic lines in uncorrelated~\cite{purrello2017} and correlated 
potentials~\cite{kolton2020thermally} in finite dimensions display logarithmic corrections that can not be accounted for by the mean-field scaling form. 

Athermal amorphous solids undergo a yielding transition well described by 
the so-called Herschel-Bulkley law relating the deformation rate 
$\dot\gamma$ and the applied stress $\sigma$

\begin{equation}
  \sigma = \sigma_c + A \dot{\gamma}^n
\end{equation}
with $n>0$ and $\sigma_c$ the critical stress; 
which sometimes is written as
\begin{equation}
  \dot{\gamma} = A^{-1}(\sigma - \sigma_c)^\beta
  \label{gamasigma}
\end{equation}
with $\beta=1/n$.
Most yield stress materials in the lab show an exponent $n$ close 
to 0.5 ($\beta \simeq 2)$, within a relatively broad range of 
variation~\cite{BonnRMP2017}.
Some of us have recently found that two-dimensional elastoplastic models 
display exponents $\beta \simeq 3/2$ or $\beta \simeq 2$, according to the local yielding rate for an over-stressed site $i$ being, respectively, constant or
stress-dependent (as $\sqrt{\sigma_i-\sigmaY_i}$, with $\sigmaY_i$ the local instability threshold)~\cite{FerreroSM2019}.
Furthermore, these rules were mapped  to the cases of ``cuspy'' and ``smooth''
disordered potentials in alternative Hamiltonian models 
for yielding~\cite{FernandezAguirrePRE2018,FerreroSM2019};
that in turn allow to understand the existence of such a $\beta$ exponent 
dichotomy when comparing them with the problem of a particle stochastically 
driven in a disordered potential~\cite{JaglaJSTAT2010},
which allows to justify those two values.

%
Along the lines followed for the depinning transition,
the zero temperature flowcurve expression (Eq.~\ref{gamasigma})
can be readily generalized to a proposal for the thermal rounding of the 
yielding transition 
\begin{equation}
\dot\gamma (\sigma,T)=  T^\psi G((\sigma-\sigma_c)/T^{1/\alpha}).
\label{eq:yielding_thermalrounding}
\end{equation}
The form of the scaling function $G$ in Eq.~(\ref{eq:yielding_thermalrounding}) 
is expected to have, for a large negative argument $x$, a leading term which is 
exponential in $x^\alpha$, reflecting in this limit the thermal activation over barriers that scale
as $(\sigma_c-\sigma)^\alpha$.
If we thus use $G(x)=C_1 \exp(C_0(-x)^\alpha)$ for large negative $x$ in 
Eq.~(\ref{eq:yielding_thermalrounding}), and invert to obtain $\sigma$,
we get
\begin{equation}
\sigma(T)=\sigma_c-(C_0^{-1}T\log(C_1T^\psi/\dot\gamma))^{1/\alpha}
\label{eq:sigmavsgammadot}
\end{equation}
which can be matched with the Johnson-Samwer expression (up to non-leading terms) 
if $\alpha$ turns out to be 3/2. 
We will see in fact that this is the value of $\alpha$ that corresponds in our simulations
to the case of smooth potentials, since they generate an energy barrier vanishing
as $(\sigma-\sigma_c)^{3/2}$ as the critical stress is approached. 

If Eq. (\ref{eq:yielding_thermalrounding}) describes correctly the full rounding of 
the transition, it must also work for positive arguments of the $G$ function, 
in particular for $T\to 0$. 
If this is the case, $G(x)$ for large positive $x$ must behave as $\sim x^{\alpha\psi}$ 
to cancel out the $T$ dependence, showing that the $\psi$ exponent is not independent
but is given in terms of the barrier exponent $\alpha$ and the flow exponent $\beta$ 
as $\psi=\beta/\alpha$.
We will show that in fact this holds for $\psi$ both in the case of ``cuspy''
($\alpha=2$) and ``smooth'' ($\alpha=3/2$) potentials, since the kind of underlying
disordered potential also determines the flowcurve exponent $\beta$.

Here we not only confirm numerically the good agreement with the scaling predicted by 
Eq.~(\ref{eq:yielding_thermalrounding}) but also clarify its origin. 
We also justify the validity of  Eq.~(\ref{eq:yielding_thermalrounding}) 
as is, without the corrections to scaling that are expected in low-dimensional
cases with short-range elasticity~\cite{kolton2020thermally}.
Therefore, as we discuss deeper in the following, the finding that for yielding 
Eq.~\ref{eq:yielding_thermalrounding} is indeed very well satisfied can be 
considered as a manifestation of the mean-field-like nature of the yielding phenomenon.

In the next section we briefly present the two main numerical approaches that we use, 
namely elastoplastic and Hamiltonian models, leaving a slightly more detailed presentation 
for Appendix~\ref{app:modeldescription}. 
Then in Section~\ref{sec:results} we show results in both kind of models, displaying a very 
robust thermal rounding scaling. 
In Section~\ref{sec:oneparticle} we interpret those results in terms of the 
single particle Prandtl-Tomlinson model in the presence of thermal and mechanical noise. 
Finally, Section~\ref{sec:discussion} contains a discussion and summary. 

\section{Models}

We run simulations of two different kind of coarse-grained models of amorphous solids:
On one hand, `Hamiltonian models' in which disorder is encoded in quenched potentials,
the evolution equation of local strains is given by forces derived from a potential and 
the stochastic process is Markovian.
On the other hand, `classical EPMs', where the instantaneous state of elastoplastic blocks 
constitutes a local `memory' and the system evolution is not necessarily Markovian.
In the following we give a minimal description of both frameworks.
See Appendix \ref{app:modeldescription} for a more complete presentation and references 
to the literature. 

\subsection{Hamiltonian Model}\label{modelHM}

In the Hamiltonian model we consider the local strain $e({\bf r},t)$, 
that we will write $e_i$ when discretized on a numerical cubic mesh. 
The temporal evolution of $e_i$ is through an overdamped dynamical equation of the form
\begin{equation}
\frac{\partial e_i}{\partial t}=- \frac{d V_i}{d e_i} +\sum_{j}G_{ij}e_j+\sigma+\sqrt {T} \xi_i(t)
\end{equation}
Here $\sigma$ is the applied stress, and the strain rate $\dot\gamma$ is calculated as 
\begin{equation}
\dot\gamma\equiv \frac{d\overline{e_i}}{dt}=-\overline{\frac{d V_i}{d e_i}}+\sigma
\end{equation} 
with the bar indicating spatial averaging.

The long range interaction term $G_{ij}$ is of the Eshelby type (see below), 
and we incorporate temperature through the stochastic term $\xi(t)$ satisfying 
\begin{eqnarray}
\langle \xi_i(t) \rangle&=&0\\
\langle \xi_i(t)\xi_j(t') \rangle&=&2\delta(t-t')\delta_{ij}.
\end{eqnarray}

The potentials $V_i$ are disordered potentials, uncorrelated in space, that represent the disordered nature of our amorphous 
material.
We consider two different forms for these potentials, that we call ``cuspy'' and ``smooth''. 
They are defined in Appendix~\ref{app:modeldescription}1.
Both cases describe potentials with many different local minima. 
The main difference between the two cases is that in the smooth case the force is continuous, 
whereas in the cuspy case there are discontinuities in the force when moving from one basin 
to the next.

\subsection{Elastoplastic Model}\label{modelEPM}

$2$-dimensional elasto-plastic models (EPMs) can be defined by a
scalar field $\sigma(\br,t)$, with $\br$ discretized on a square 
lattice and each block $\sigma_i$ subject to the following evolution 
in real space
\begin{equation}\label{eq:eqofmotion1}
\frac{\partial \sigma_i(t)}{\partial t} =
  \mu\dot{\gamma}^{\tt ext}  +\sum_{j} G_{ij} n_j(t)\frac{\sigma_j(t)}{\tau} ;
\end{equation}
where $\dot{\gamma}^{\tt ext}$ is the externally applied strain rate, 
and the kernel $G_{ij}$ is the Eshelby stress propagator~\cite{Picard2004}.
$G_{ii} < 0$ sets the local stress dissipation rate for an active site.
The form of $G$ is $G(\br,\br') \equiv G(r,\varphi)\sim\frac{1}{\pi r^2}\cos(4\varphi)$ in polar coordinates,
where $\varphi \equiv \arccos((\br-\br')\cdot\br_{\dot{\gamma}^{\tt (ext)}})$ and
$r \equiv \left|\br-\br'\right|$. For our simulations we obtain $G_{ij}$ from the values of the propagator in Fourier space $G_{\bf q}$, defined as
\begin{equation}
G_{\bf q} = -\frac{4q_x^2q_y^2}{(q_x^2+q_y^2)^2}.
\label{eshelby_kernel}
\end{equation}
for $\bf q\ne 0$, and $G_{\bf q=0}=-1$.

The elastic shear modulus $\mu=1$ defines the stress unit, and the 
mechanical relaxation time $\tau=1$, the time unit of the problem.
The last term of (\ref{eq:eqofmotion1}) (for $j\neq i$) constitutes a 
\textit{mechanical noise} acting on $\sigma_i$ due to the instantaneous 
integrated plastic activity over all other blocks in the system.

The picture is completed by a dynamical law for the local state variable 
$n_i=\{0,1\}$.
Typically a block yields ($n:0\to 1$) when its stress $\sigma_i$ reaches 
a threshold $\sigmaY_i$ and recovers its elastic state ($n:1\to 0$)
after a stochastic time of order $\tau$. 
The important addition to classic models is that we now allow for thermal 
activation at finite temperature $T>0$.
This is, even when the local stress is below the local threshold
$\sigma_i<\sigmaY_i$, the site is activated with probability 
$\exp(-(\sigmaY_i-\sigma_i)^\alpha/T)$ per unit time.
See Appendix~\ref{app:modeldescription} for all the details.

\section{Results}
\label{sec:results}

\subsection{Thermal rounding in the Hamiltonian Model}

\begin{figure}[t!]
\includegraphics[width=\columnwidth]{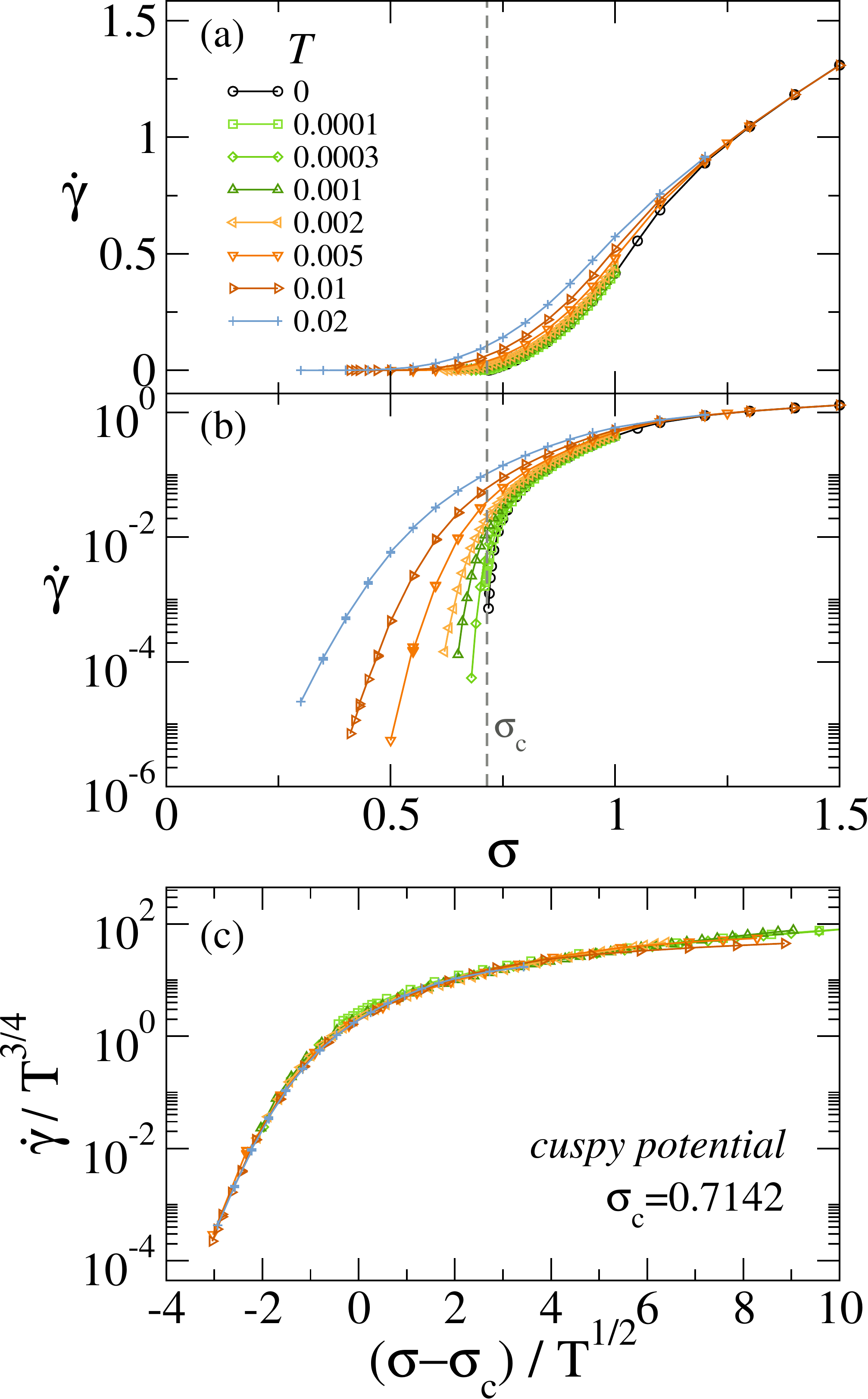}
\caption{
(a) Different temperature flowcurves for a two-dimensional Hamiltonian system
of size 512 $\times$ 512, with parabolic potentials. 
The rightmost curve, $T=0$, fits well a form $\dot\gamma\sim (\sigma-\sigma_c)^{3/2}$
close to $\sigma=\sigma_c=0.7142$ (indicated by a vertical dashed line).
(b) Same data in log-lin scale. 
(c) Scaling using $\psi=3/4$, $\alpha=2$ (compatible with $\beta=\psi\alpha=3/2$).
}
\label{fig:512parab}
\end{figure}

\begin{figure}[t!]
\includegraphics[width=\columnwidth]{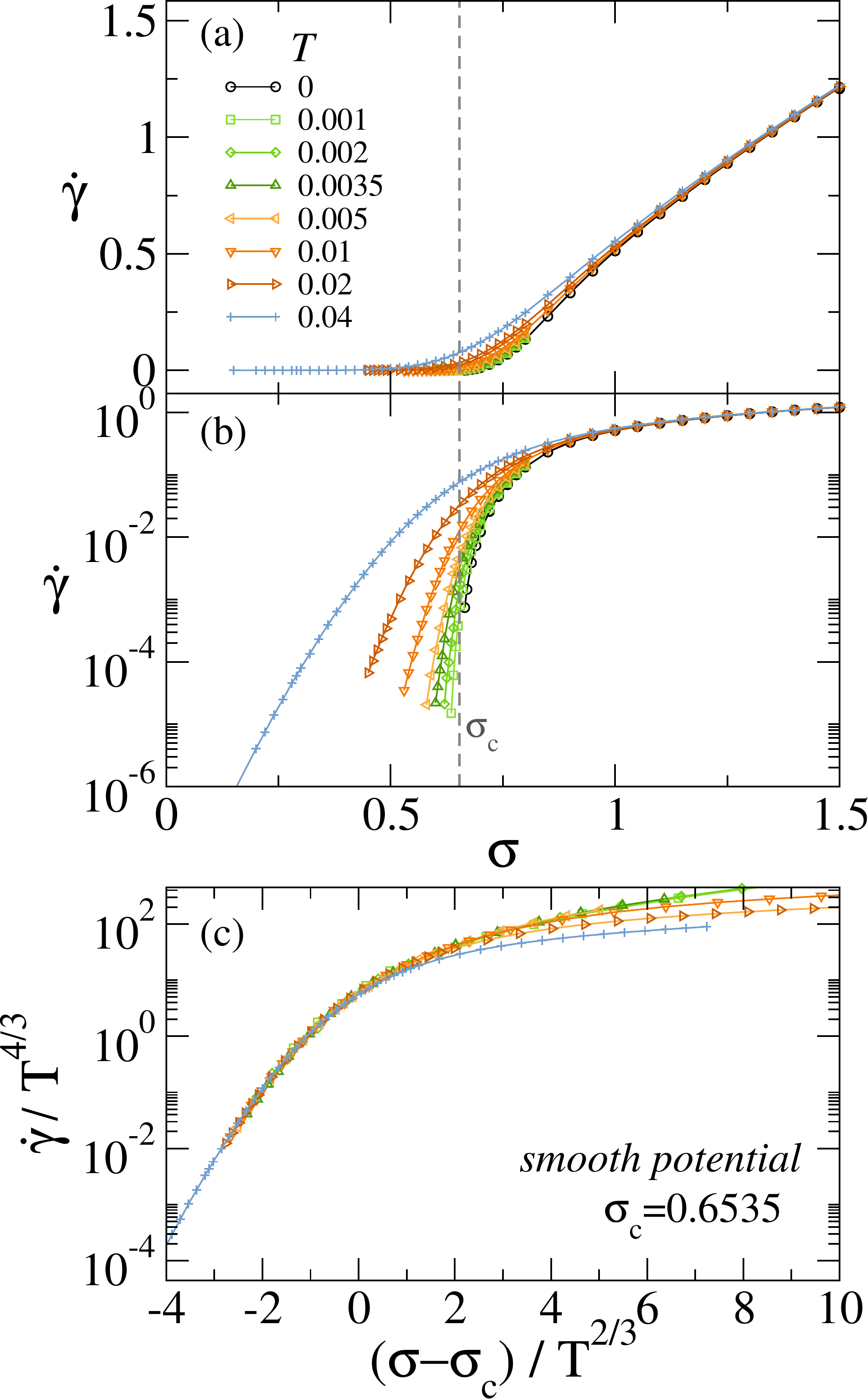}
\caption{
(a) Flowcurves at different temperatures for a two-dimensional Hamiltonian 
system of size 256 $\times$ 256, with smooth potentials. 
The rightmost curve, $T=0$, fits well a form $\dot\gamma\sim (\sigma-\sigma_c)^{2}$
close to $\sigma=\sigma_c=0.6535$ (indicated by a vertical dashed line). 
(b) Same data in log-lin scale. 
(c) Scaling using $\psi=4/3$, $\alpha=3/2$ (compatible with $\beta=\psi\alpha=2$).
}
\label{fig:256smooth}
\end{figure}

We start by presenting the results of our numerical simulations
for the Hamiltonian model described in Sec.~\ref{modelHM}.
In Fig.~\ref{fig:512parab}(a) we see the strain rate vs. stress curve 
(the flow-curve) in a 2D system of size 512$\times$ 512, with 
disordered potentials formed by concatenation of parabolas which is 
an instance of the ``cuspy'' case. 
The effect of temperature is clearly visible as it generates finite 
values of strain rates even below the zero temperature critical 
stress $\sigma_c$, that was estimated from a power law fit of the 
flowcurve corresponding to zero temperature.
From the zero temperature fitting we also determine the flow exponent, that 
to a good approximation turns out to be $\beta=3/2$, as previously reported 
for this kind of potential~\cite{FernandezAguirrePRE2018}.
The effect of temperature is more clearly visible plotting the $y$ axis 
in logarithmic scale (Fig.~\ref{fig:512parab}(b)). 
In this logarithmic plot, it is also clearer the possibility to collapse 
curves at different temperatures in a single scaled curve.
This is done in Fig.~\ref{fig:512parab}(c), using the scaling proposed in Eq. 
(\ref{eq:yielding_thermalrounding}) with $\alpha=2$ and $\psi=3/4$.
The scaling collapse is very good. 
It extends to a wide range around $\sigma_c$ (at least of $\simeq 30 \%$ of $\sigma_c$), 
and to temperatures up to $\simeq 0.02$ (to be compared with the reference energy 
value $\sim 1$ that is the typical height of local barriers when $\sigma=0$).

The rational behind the values of the exponents used in the previous
scaling collapse is the following. 
The value of $\alpha=2$ indicates that the activation barrier 
for $\sigma$ slightly below $\sigma_c$ grows as $(\sigma_c-\sigma)^2$, 
which is consistent with the straightforward result obtained in a 
one-particle system (see Sec.~\ref{sec:oneparticle}), considering 
the cusps between successive parabolic pieces of the potential
in which a particle moves. 
As we know from previous works~\cite{jagla2018prandtl,FernandezAguirrePRE2018,FerreroSM2019}, 
the kind of disordered potential will also determine $\beta$.
Therefore we say that $\alpha$ and $\beta$ should be `compatible'.
In fact, if Eq.~\ref{eq:yielding_thermalrounding} is to be applicable to the 
limit $T\to 0$, then the $T$ 
dependence in this limit must vanish, providing $\psi=\beta /\alpha=3/4$ ($\beta=3/2$ 
and $\alpha=2$ in this case), which is precisely the value used in 
Fig.~\ref{fig:512parab}(c) to obtain a good collapse of the data.

A similar analysis and scaling can be done for the case of smooth potentials,
here constructed by combining sinusoidal functions (see Appendix~\ref{app:modeldescription}).
Results (for a system of size $256 \times 256$) are shown in Fig.~\ref{fig:256smooth}. 
We see that in this case the range of validity of the scaling is somewhat more 
limited in extent than in the previous case.
This is simply a consequence of the fact that the extent of the critical region of the $T=0$ 
case is smaller~\footnote{At large enough values of $\sigma$ the system will always crossover 
to a fast-flow regime where $\dot\gamma \sim \sigma$.}.
The values of the exponents that are expected to fulfill the scaling are
$\beta=2$ and $\alpha=3/2$ (see Sec.~\ref{sec:oneparticle} and \cite{jagla2018prandtl, FernandezAguirrePRE2018}).
Requiring the exponent $\psi$ to satisfy the relation $\psi=\beta/\alpha$, it results $\psi=4/3$.
From the collapse of Fig.~\ref{fig:256smooth}(c) we conclude that the scaling of
Eq.~\ref{eq:yielding_thermalrounding} works perfectly well also in the present case of
$\beta=2$ and $\alpha=3/2$, thus indicating that the thermal rounding scaling is robust 
with respect to details of the form of the disordered potential.

\subsection{Thermal rounding in elasto-plastic models}

\begin{figure}[t!]
\includegraphics[width=\columnwidth]{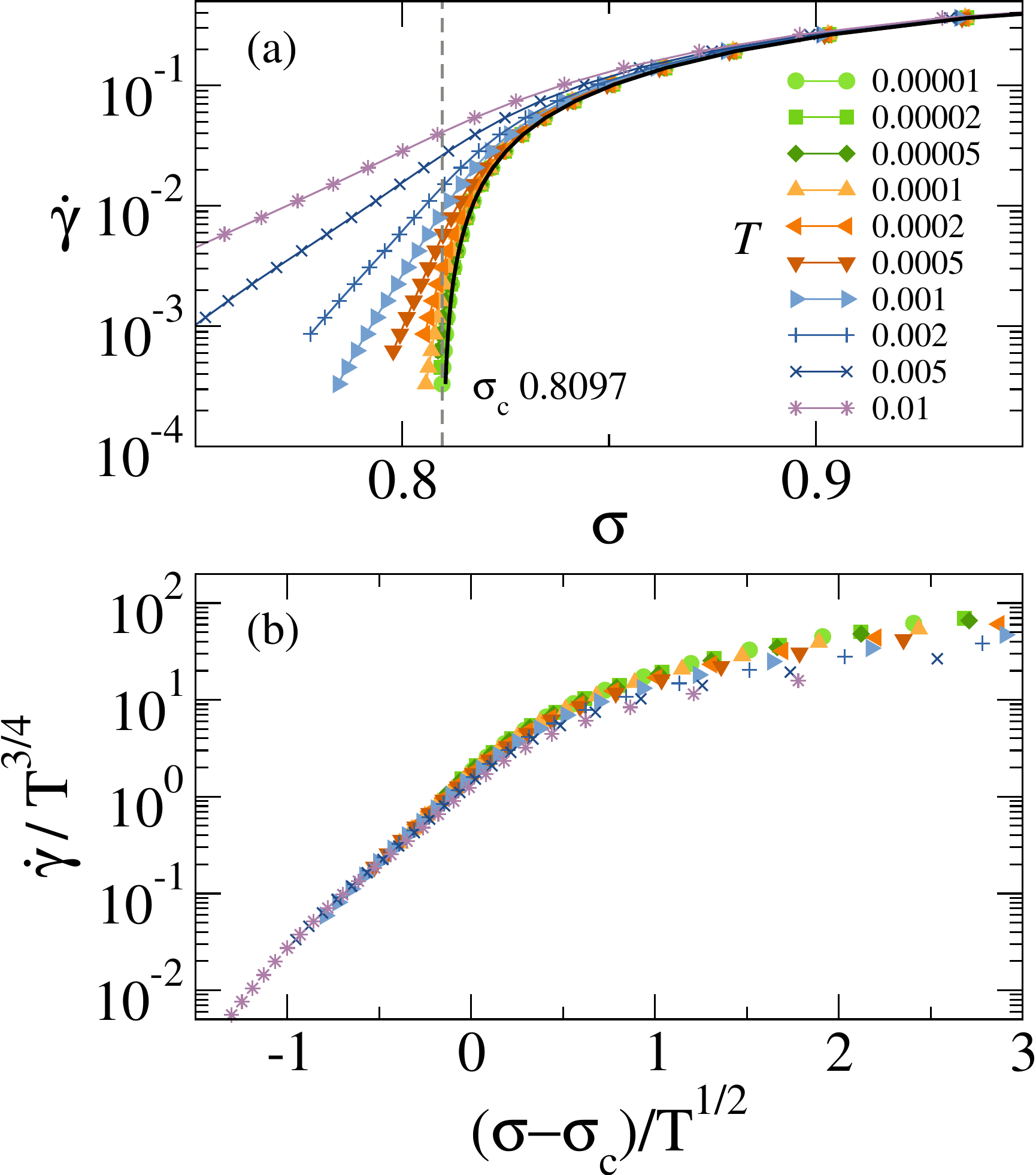}
\caption{
(a) Flowcurves at different temperatures for a two-dimensional 
EP model of size $2048 \times 2048$, with renewable thresholds taken 
randomly from a distribution and instantaneous plastic events.
The barrier coefficient used is $\alpha=2$.
The continuous black line corresponds to $T=0$.
(b) Scaling~(\ref{eq:yielding_thermalrounding}) using $\psi=3/4$, $\alpha=2$ 
(compatible with $\beta=\psi\alpha=3/2$).
}
\label{fig:epm_instantaneaous_exprandomthresholds}
\end{figure}

We now test the thermal rounding scaling of Eq.~\ref{eq:yielding_thermalrounding}
in different EPMs.
First, notice that many classical EPMs (e.g. \cite{Picard2004,LinEPL2014, LinPNAS2014})
consider a common local threshold for all sites and a local stochastic rule to define 
the precise moment of the local yielding.
In the construction of $T>0$ EPMs we have chosen instead to use distributed local
thresholds (as in ~\cite{nicolas2014rheology}) and immediate yielding upon reaching 
the threshold, avoiding an extra stochastic rule for the site activation.
Instead, we now include the possibility for a site to be activated by temperature,
with a probability $\exp\left[-(\sigmaY_i-\sigma_i)^\alpha/T\right]$ (see Appendix \ref{app:modeldescription}).

In Fig.~\ref{fig:epm_instantaneaous_exprandomthresholds} we show flowcurves
at different temperatures for an EPM with exponentially distributed thresholds
($\sigmaY_i = 1+0.1 r_e$, with $r_e$ an exponentially distributed random number)
and \emph{instantaneous} plastic events, i.e., the stress relaxation occurs in 
a single time step.
Panel (a) shows the flowcurves in log-lin scale. 
Using a $\sigma_c=0.8095$ obtained by extrapolating the $T=0$ flowcurve, 
the expected ``universal'' exponent $\beta=3/2$ for this kind of 
EPM~\footnote{EPMs with a \emph{uniform} yielding rate are analogous
to the case of `cuspy' potentials~\cite{FerreroSM2019}.},
and the corresponding value of $\alpha=2$ used in the activation rule,
we observe a good collapse in a wide range of temperatures and strain 
rates for the scaling~(\ref{eq:yielding_thermalrounding}) with $\psi=\beta /\alpha=3/4$.

\begin{figure}[t!]
\includegraphics[width=\columnwidth]{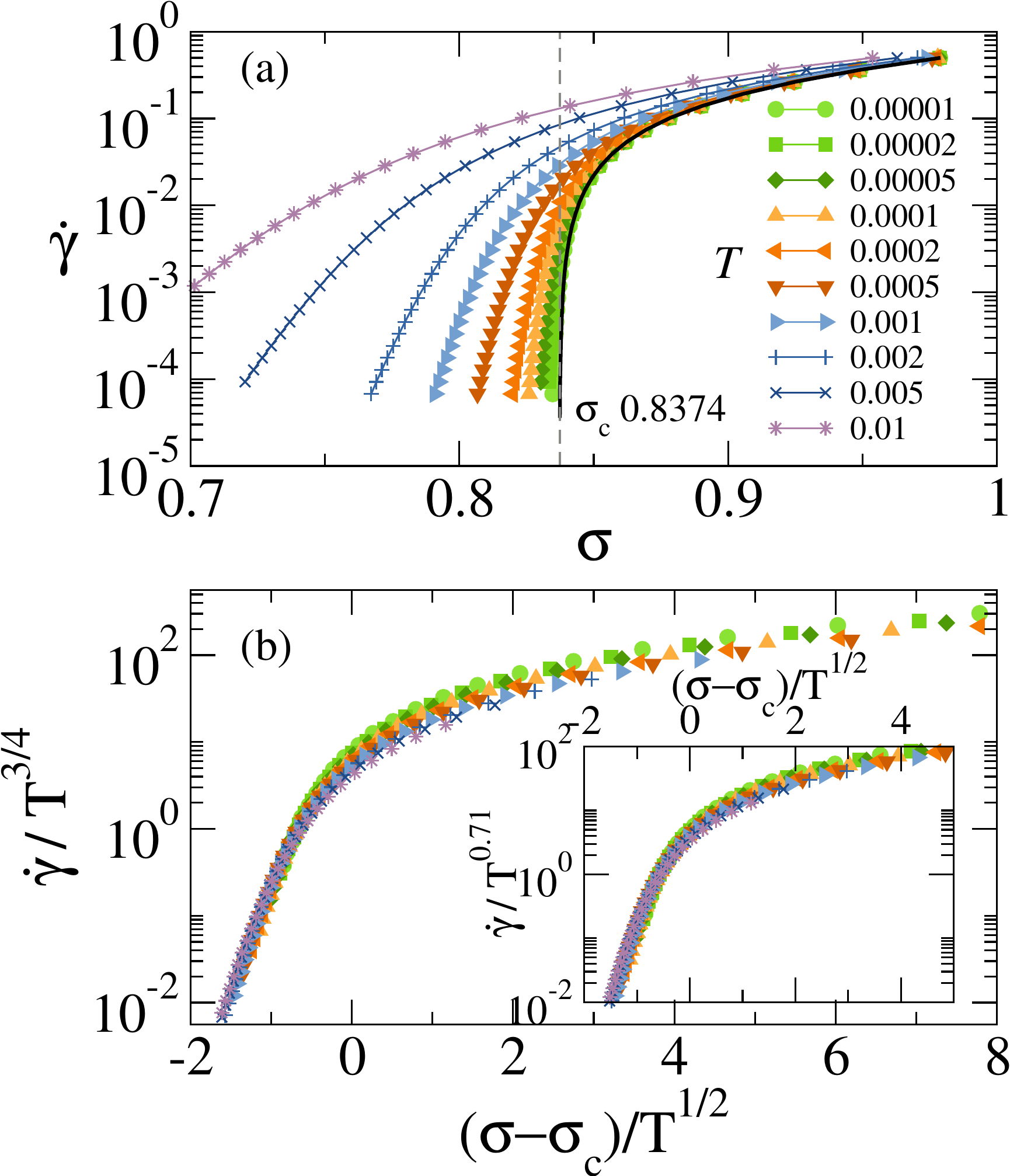}
\caption{
(a) Flowcurves at different temperatures for a two-dimensional 
EP model of size 2048 $\times$ 2048, with renewable thresholds taken 
randomly from a distribution and finite duration plastic events.
The barrier coefficient used is $\alpha=2$.
The continuous black line corresponds to $T=0$.
(b) Main plot: scaling~(\ref{eq:yielding_thermalrounding}) using $\psi=3/4$, $\alpha=2$ 
(compatible with $\beta=\psi\alpha=3/2$)
Inset: scaling~(\ref{eq:yielding_thermalrounding}) but with an effective value 
$\beta=1.42$ used instead, keeping $\alpha=2$ and $\psi=\beta/\alpha$.
}
\label{fig:epm_uniformthresholds_finiteduration}
\end{figure}

If we now add a bit more of phenomenology in the elastoplastic modeling
and  allow for a `finite duration' of plastic 
events~\footnote{Of physical relevance even for overdamped systems where it is 
expected to scale as the ratio between an effective microscopic viscosity and 
the elastic shear modulus~\cite{NicolasRMP2018}.},
we start to loose the formal analogy with the Hamiltonian systems.
In particular, the dynamics is now non-Markovian due to the local state memory
(see~(\ref{app_eq:activationrulesEPM})).
Yet, we can still test the thermal scaling~\ref{eq:yielding_thermalrounding}.
In fact, flowcurves at $T=0$ for EPMs with finite event duration were
seen to recover the $\beta$ exponent prescribed by the Hamiltonian systems
and derived from the PT model, but only at small enough strain-rate values
and with deviations out of the scaling regime ascribed precisely to the
finite duration of the events~\cite{FerreroSM2019}.

Figure \ref{fig:epm_uniformthresholds_finiteduration} shows flowcurves at 
different temperatures for an EPM with randomly distributed thresholds 
($\sigmaY_i = 1+0.1 r_e$, with $r_e$ an exponentially distributed random number)
and a finite duration for plastic events ($\tau_{\tt off}=1$).
We first observe that the finite duration of the events modifies
the estimated critical stress $\sigma_c$, which is highly non-universal,  
respect to the one of Fig.~\ref{fig:epm_instantaneaous_exprandomthresholds}.
The scaling displayed in the main plot of panel (b), using $\psi=3/4$ and 
$\alpha=2$ is not bad, but also not perfect.
In paticular, it is displeasing to see that the curves do not strictly collapse
at $\sigma=\sigma_c$.
Since the true critical region and the existence of universal exponents can be limited to
a range of very small strain rates and stresses around the critical point, 
we checked the possibility of having a better collapse with an effective value of $\beta$ that takes into account
the possibility of corrections to the ideal scaling.
The inset of Fig.~\ref{fig:epm_uniformthresholds_finiteduration}(b) uses 
$\beta\simeq 1.42$ while keeping $\alpha=2$ and $\psi=\beta/\alpha(=0.71)$.
This scaling looks better in a wider range of $\dot\gamma - \sigma$ values and
the scaling assumption~(\ref{eq:yielding_thermalrounding}) perfectly holds.
The effective exponent $\beta\simeq 1.42$ is the one that we would fit from the 
flowcurve at $T=0$ in the range of stresses $[1\times 10^{-4} , 5\times 10^{-3}]$ above $\sigma_c$.

Previous works~\cite{jagla2018prandtl,FernandezAguirrePRE2018,FerreroSM2019,FerreroPRL2019}
have indicated that EPMs and the Hamiltonian description are equivalent in some limiting cases. 
The typical EP modeling (that considers a constant activation probability once a local stress 
threshold is overpassed) has been seen to correspond to a Hamiltonian model that uses a 
``cuspy'' form for the pinning potential. In order to represent the case of ``smooth'' potentials
EPMs have to use a ``progressive'' activation law, as described in ~\cite{FerreroSM2019}.
This analogy reflects qualitatively the way in which a block escapes from a local solid state and 
moves to the next one by local fluidization and the typical time it takes to do so~\cite{FerreroSM2019}.
The matching is further reinforced by the finding that the flow exponent $\beta$ 
is very close to 3/2 both in Hamiltonian models with cuspy potentials and EPMs with 
uniform activation, whereas $\beta\simeq 2$ is found in Hamiltonian models with 
smooth potentials and EPMs with the appropriate progressive activation. 
Therefore, we believe that once the kind of barrier has been selected (equivalently,
the type of local yielding rule), both $\alpha$ and $\beta$ are simultaneously defined,
i.e., they are not independent exponents in any physically relevant situation.

Recently Popovic et al.~\cite{popovic2020thermally} have presented a study of thermal 
rounding in elastoplastic models, finding a very good scaling of the form of 
Eq. (\ref{eq:yielding_thermalrounding}).
While they have varied $\alpha$ freely to nicely test the scaling for different kind 
of thermal activations, the article unfortunately does not discuss in detail the 
values of $\beta$ (and therefore $\psi$) used for the scaling. 
Furthermore, there is no discussion on the ``curious'' fact that the scaling obtained 
is extremely good, actually more than expected in other cases of thermal rounding of 
models with non mean-field scaling~\cite{kolton2020thermally}.
Interestingly, one could interpret that such a good mean-field-like scaling is somehow 
in contrast with previous expectations from the same group about the yielding exponent 
$\beta$ being non-universal and significantly affected by finite-dimensional 
effects~\cite{LinPRE2018}.
We think that the excellent performance of the thermal rounding scaling (\ref{eq:yielding_thermalrounding})
is not a fortuitous coincidence, but instead
a consequence of the fact that the yielding transition is effectively mean-field, 
as we discuss in the next Section.

\begin{figure}[t!]
\includegraphics[width=\columnwidth]{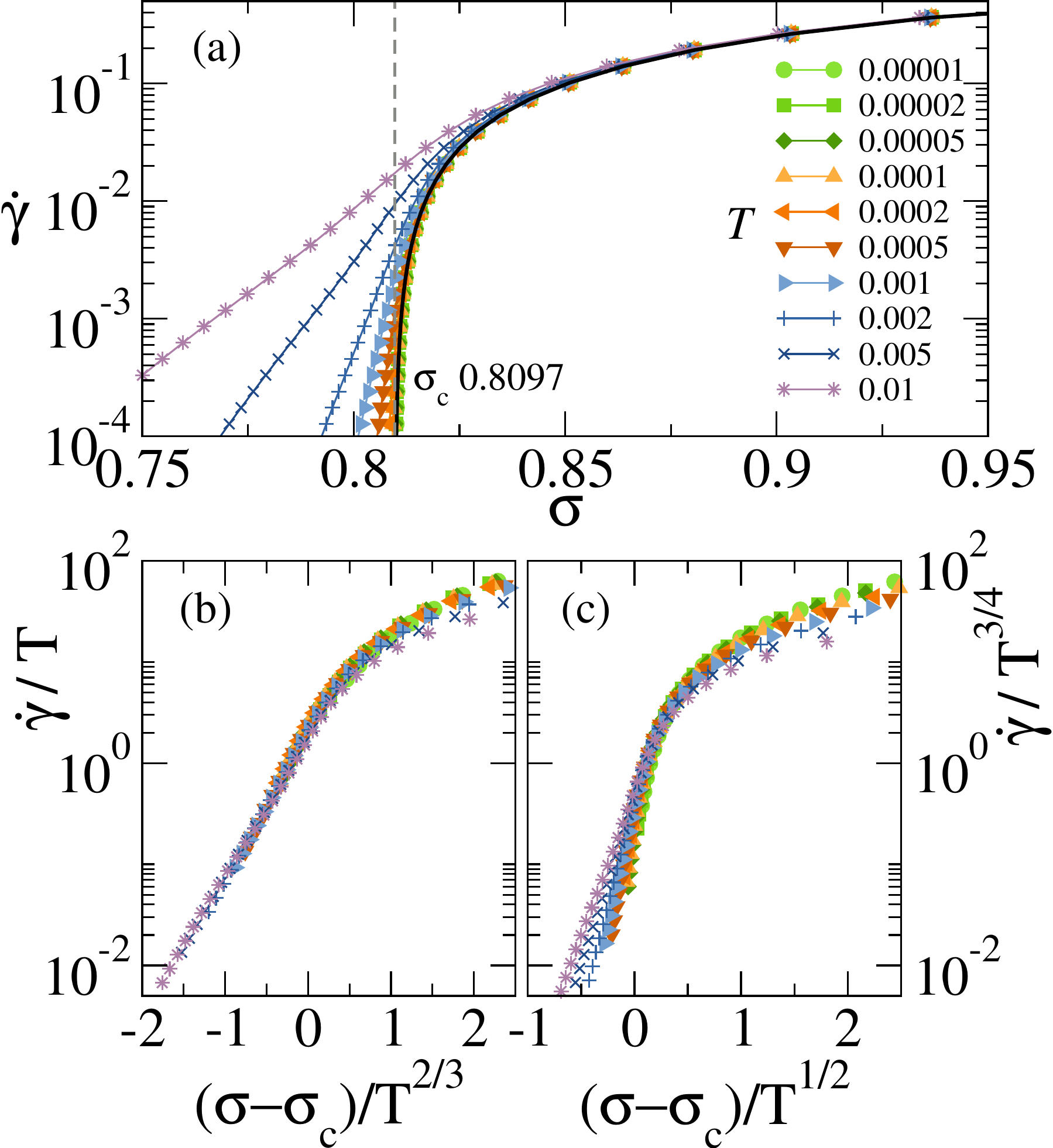}
\caption{
(a) Different temperature flowcurves (in log-lin) for a two-dimensional 
EP model of size 2048 $\times$ 2048, with thresholds taken randomly from a 
distribution, renewed after each plastic event, and instantaneous plastic events.
The barrier coefficient used is $\alpha=1.5$.
The continuous black line corresponds to $T=0$.
(b) Scaling~(\ref{eq:yielding_thermalrounding}) using $\psi=1$, $\alpha=3/2$ 
(compatible with $\beta=\psi\alpha=3/2$).
(c) Wrong scaling, using $\psi=3/4$, $\alpha=2$ 
(compatible with $\beta=\psi\alpha=3/2$).
}
\label{fig:epm_uniformthresholds_instantaneous_alpha1p5}
\end{figure}

As a matter of fact, one way to test the mean-field-like hypothesis is to build
a system in which we define arbitrarily the kind of activation barrier under consideration.
This may lead to an equally arbitrary value of $\psi$.
But, as soon as we know that the characteristics that determines $\beta$ is preserved,
we can expect the same thermal rounding scaling to hold, with $\psi=\beta/\alpha$.

In Fig.~\ref{fig:epm_uniformthresholds_instantaneous_alpha1p5} we use an 
``incorrect'' value of alpha in the activation barrier, $\alpha=1.5$, while
the plastic events still occurs at a fix rate (as soon as they reach the local threshold).
Therefore, with the $\beta=3/2$ expected for constant rates, the scaling~(\ref{eq:yielding_thermalrounding}) translates to 
$\dot\gamma/T$ vs. $(\sigma-\sigma_c)/T^{2/3}$.
This is what is plotted in Fig.~\ref{fig:epm_uniformthresholds_instantaneous_alpha1p5}(b).
Despite small deviations to scaling away from $\sigma_c$ for the higher temperatures, 
which are expected, the scaling behaves quite well.
Yet, notice that the scaling exponents that worked well 
in Figs.~\ref{fig:epm_instantaneaous_exprandomthresholds} and~\ref{fig:epm_uniformthresholds_finiteduration}, 
i.e., $\dot\gamma/T^{3/4}$ vs. $(\sigma-\sigma_c)/T^{1/2}$, now
completely fail, as is shown in 
(Fig.~\ref{fig:epm_uniformthresholds_instantaneous_alpha1p5}(c)).
So, even when we can link $\beta$ at $T=0$ with the type of local yielding
rule (constant or progressive when reaching threshold), if we mix that rule 
with a thermal activation governed by $\alpha$, the scaling
relation is still expected to be Eq.~\ref{eq:yielding_thermalrounding}
with $\psi=\beta/\alpha$.
This is why we believe that the scaling works so well for all $\alpha$
in Ref.~\cite{popovic2020thermally}, even when $\beta$ should be similar
in all cases, and therefore $\psi$ should be changing.

In brief, we observe that the thermal rounding of Eq.~(\ref{eq:yielding_thermalrounding})
works well in elastoplastic models where the possibility of thermal activation
has been introduced in an Arrhenius-like fashion ($\sim\exp(-(\sigmaY_i-\sigma_i)^\alpha/T)$).
While the EPMs results alone could leave space for interpretation due to the 
effective $\beta$ exponents measured, the analogy with the Hamiltonian models
strongly suggests that, in the background, the thermal rounding scaling is 
working with no corrections.
This places the yielding phenomenon, beyond the athermal limit, on the 
spot of a mean-field-like or particle-based theoretical 
interpretation~\cite{FerreroSM2019,FerreroPRL2019,FerreroJPCM2021}, provided that
the non-trivial mechanical noise is well characterized for each dimension.
In the next section we combine thermal and mechanical noises in such a 
one-particle problem.

\section{One particle under mechanical and thermal noise}
\label{sec:oneparticle}

The finding that our numerical results accurately follow the scaling 
of Eq.(\ref{eq:yielding_thermalrounding}) with fully consistent values of the 
exponents provides additional support to a developing 
idea~\cite{FernandezAguirrePRE2018,jagla2018prandtl,FerreroSM2019,FerreroPRL2019}: 
The yielding transition in finite dimensions can be accurately described by a 
mean-field-like model in which a single site feels the effect of all other sites through 
a ``mechanical noise'' characterized by a time signal with a non-trivial 
Hurst exponent $H$~\cite{FerreroPRL2019}. 
In Ref.\cite{jagla2018prandtl} this one-particle model was analyzed in detail at zero 
temperature, and it was shown that the value of the flow exponent $\beta$ is related to 
the value of $H$ by
\begin{equation}
  \beta=\begin{cases}
    \frac 1H, & \text{``cuspy potential''}, \\
    \frac 1H+\frac 12, & \text{``smooth potential''}.
  \end{cases}
  \label{eq:betapt}
\end{equation}
where the potentials are periodic and equivalent to the onsite potentials 
defined for the 2D systems of Sec.\ref{modelHM} (Eq.~\eq{eq:potentials2d}),
\begin{equation}
  -\frac{dV}{dx}=\begin{cases}
    [x]-x, & \text{``cuspy''},\\
    \sin(2 \pi x), & \text{``smooth''}.
  \end{cases}
  \label{eq:potentialsone}
\end{equation}

Taking into account that $\beta=3/2$ and $\beta=2$ for the cuspy and smooth potentials, 
respectively, we see that $H$ in the one particle effective model must be taken 
as~\cite{jagla2018prandtl} $H=2/3$. 
With this choice, we will show in the following that the addition of an additive thermal 
noise leads to a good agreement both with the overall form of Eq.(\ref{eq:yielding_thermalrounding}), 
and also with the numerical values of the thermal rounding exponents found in the previous 
sections for the two-dimensional models.
Notice that the choice of periodic potentials is done for convenience, 
since it allows for more straightforward analytic approximations, 
but the use of bounded disordered potentials keeping the same `cuspy' or `smooth' 
characteristics would yield identical results. 

\begin{figure}[t!]
\includegraphics[width=\columnwidth]{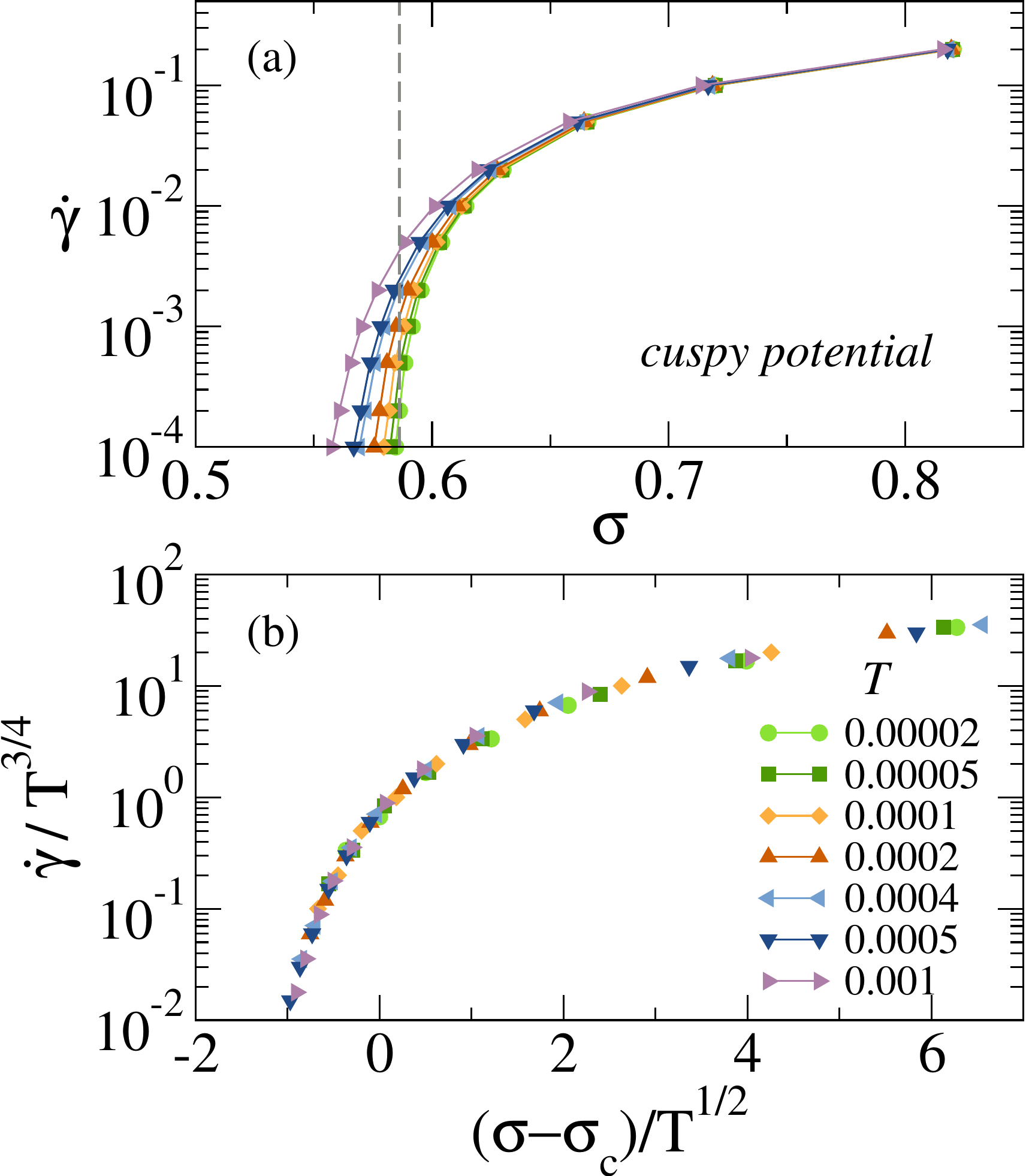}
\caption{
(a) Flow-stress curves for one particle in a cuspy periodic 
potential composed by concatenated parabolic wells, stochastically 
driven with mechanical noise of $H=2/3$ at different temperatures. 
(b) Master curve (Eq.\ref{eq:yielding_thermalrounding})
using the exponents corresponding to the $\omega=1$ case, $\psi=3/4$ 
and $1/\alpha=1/2$ from Eq.(\ref{eq:exponentsPT}), and
$\sigma_c=0.5862$. 
}
\label{fig:1pla1}
\end{figure}

\begin{figure}[t!]
\includegraphics[width=\columnwidth]{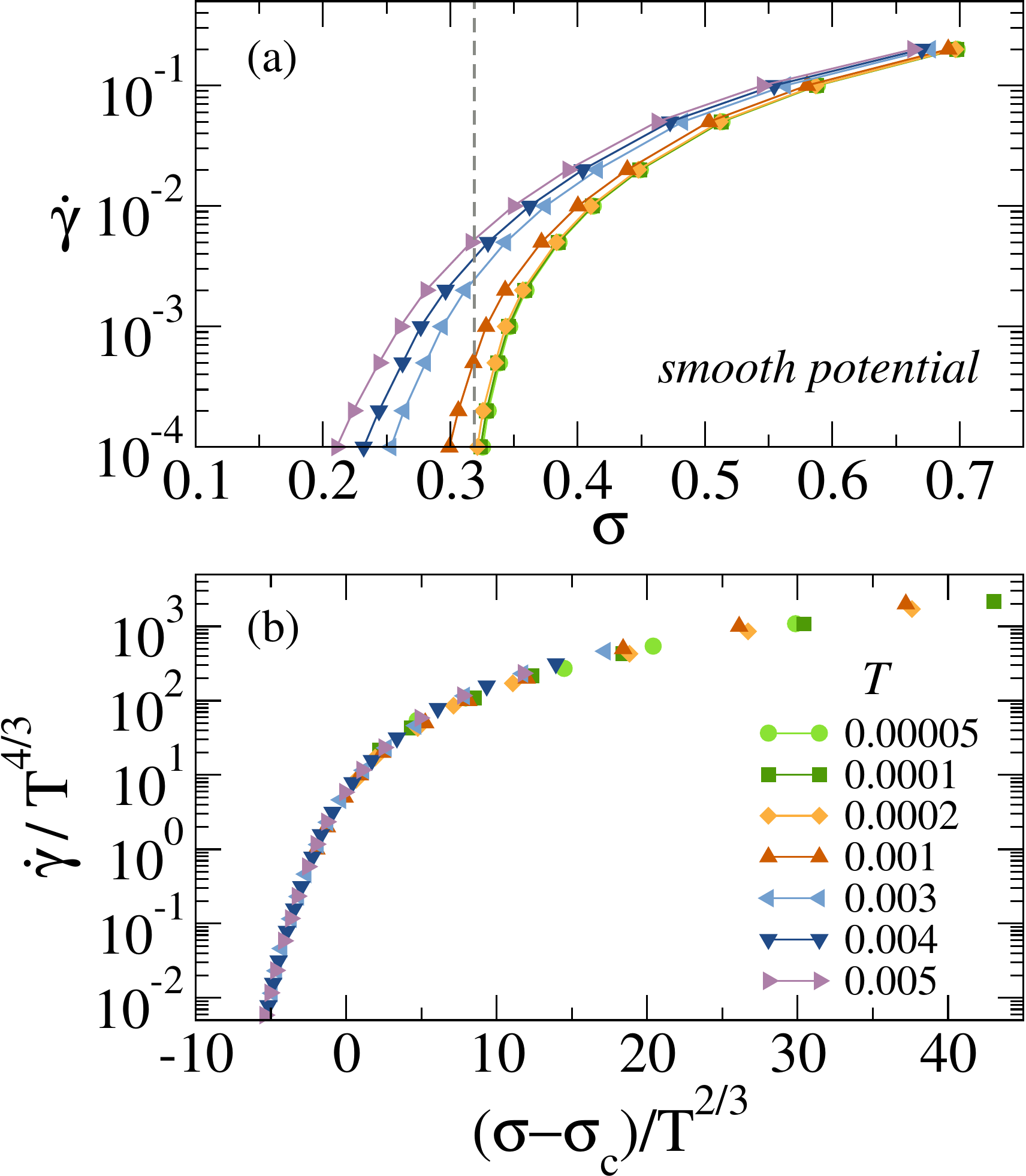}
\caption{
(a) Flow-stress curves for one particle in a sinusoidal potential, 
stochastically driven with mechanical noise of $H=2/3$ at different temperatures. 
(a) Master curve (Eq.\ref{eq:yielding_thermalrounding})
using the exponents corresponding to the $\omega=2$ case, $\psi=4/3$ 
and $1/\alpha=2/3$ from Eq.(\ref{eq:exponentsPT}).
$\sigma_c=0.319$.
}
\label{fig:1pla2}
\end{figure}

The model that we now simulate consists of a particle with a single 
coordinate $x$, evolving in a potential $V(x)$, driven by the 
position variable $w(t)$ through a spring of constant $k_0$, and
in the presence of a stochastic term that takes into account the 
effect of temperature $T$ in the system,
\begin{equation}
\frac{dx}{dt}=-\frac{dV}{dx}+k_0\left(w(t)-x\right) +\sqrt{T}\eta_0(t)
\label{eq_1pla1}
\end{equation}
where $\eta_0$ is taken as an uncorrelated Gaussian variable such that 
$\langle \eta_0(t)\eta_0(t')\rangle=2\delta(t-t')$. 
Hence, if $w$ is fixed, the system spontaneously relax to the Boltzmann 
equilibrium distribution in the potential $V(x)+k_0(w-x)^2/2$.

The dynamics of the variable $w(t)$ has a smooth part, 
that mimics the uniform external driving, and a stochastic 
term that represents the existence of mechanical noise in the system, 
\begin{equation}
\frac{dw}{dt}=b\dot \gamma+a{\dot\gamma}^H \eta_H(t).
\label{eq_1pla2}
\end{equation}
The mechanical noise term is characterized by the Hurst exponent $H$. 
To implement it, we sample a random variable with a heavy tailed 
probability distribution, 
\begin{equation}
P(\eta_H)\sim \frac 1{|\eta_H|^{{\frac 1H}+1}}, ~~~~\mbox {for large}~~|\eta_H|.
\label{eq_1pla3}
\end{equation}
In practice, we sample it as 
\begin{equation}\label{eq:etaH}
\eta_H=S
\left(\frac{1}{R^H+\epsilon}-1\right),
\end{equation}
where $R$ is a flat random variable between 0 and 1, $S=\pm 1$ is a binary 
random variable satisfying $\langle S_t S_{t'}\rangle=\delta(t-t')$ and 
$\langle R_t R_{t'}\rangle=\delta(t-t')/3$, and $\langle S_t R_{t'}\rangle=0$. 
Hence, $\eta_H$ is also time-decorrelated.
It is easy to see that this sampling generates Eq.(\ref{eq_1pla3}) with a 
large $\eta_H$ cut-off controlled by $\epsilon$.

We have numerically solved the stochastic system of Eqs.(\ref{eq_1pla1}) and (\ref{eq_1pla2}) 
for different values of $T$ and $\gdot>0$ in order to obtain the flow-stress 
($\gdot$ vs $\sigma$) curves near the yielding transition, with the stress 
$\sigma$ given by the steady-state average
\begin{equation}
\sigma(\gdot) \equiv k_0 (\overline{w(t)-x(t)}).
\end{equation}
Without loss of generality, in simulations we used the values $k_0=0.2$ or $k_0=0.5$. 
These values satisfy the condition $k_0 < \max_x [-V''(x)]$ for the cuspy and 
smooth potentials, thus granting $\sigma(\gdot~\to~0) = \sigma_c>0$.
We set $\epsilon$ small enough so to assure that scaling exponents
are independent of $k_0$ and $\epsilon$~\cite{jagla2018prandtl}. 

The numerical data for the flowcurves shown in Figs. \ref{fig:1pla1} and \ref{fig:1pla2}
show a very good qualitative agreement with the results found for the full 2D system, 
both for the ``cuspy'' and ``smooth'' cases of Eq.(\ref{eq:potentialsone}). 
In particular, Fig.~\ref{fig:1pla1}(b) and Fig.~\ref{fig:1pla2}(b) show 
a very good scaling collapse when using the expected values $\alpha=2$ and 
$\beta=3/2$ (and then $\psi=\beta/\alpha=3/4$) for the $\omega=1$ ``cuspy'' case, 
and $\alpha=3/2$, $\beta=2$ (and then $\psi=\beta/\alpha=4/3$) for the $\omega=2$ 
``smooth'' case. 
In fact, it can be analytically shown (see Appendix~\ref{sec:appendixoneparticle}) 
that for low enough temperatures, and sufficiently close to the critical stress $\sigma_c$, 
the flowcurves at different temperatures for the one-particle problem can be cast in 
the scaled form given by Eq. (\ref{eq:yielding_thermalrounding}), with the values of the 
scaling exponents
\begin{eqnarray}
\psi&=&\frac{\omega-H+\omega H}{(\omega+1)H}\\
\alpha&=&1+\frac{1}{\omega}\\
\beta&=&\psi \alpha=\frac 1H-\frac 1\omega +1
\label{eq:exponentsPT}
\end{eqnarray}
Here $\omega$ is related to the form of the potential $V(x)$ in Eq.(\ref{eq_1pla1}) 
right at the transition point between successive wells: 
$\omega=1$ for the ``cuspy'', and $\omega=2$ for the ``smooth'' potentials  defined 
in Eq.(\ref{eq:potentialsone}) (see Appendix~\ref{sec:appendixoneparticle} 
for a realization of $V$ with a generic $\omega$ and its thermal rounding scaling).
It is worth noting from Eq.~\eq{eq:exponentsPT} that the exponents predicted 
are universal, in the sense that they do not depend on the particular shape 
of $V(x)$ but only on its normal form near the local yielding thresholds. 

\section{Interpolation between activated and athermal flowcurves. Generalizing the Johnson and Samwer's law}

If in Eq. (\ref{eq:sigmavsgammadot}) we set $\gdot=\gdot_0$, where $\gdot_0$ is such that strain-rates below it are experimentally undetectable, 
then Eq. (\ref{eq:sigmavsgammadot}) can be considered to be a restating of the 
J\&S result, Eq. (\ref{eq:JandS}). 
In other words, the empirical finite-temperature yield stress $\tau_{c,T}$ 
appearing in Eq.\eq{eq:JandS} can be identified with the stress evaluated at the threshold strain-rate, $\tau_{c,T}\approx\sigma(\gdot_0, T)<\sigma_c$ ~\footnote{Notice 
that from this viewpoint, it is therefore clear that the J\&S scaling 
corresponds to the thermally activated regime, $\sigma<\sigma_c$.}.
More generally, we can propose an interpolation scheme between the exponentially activated regime at $\sigma<\sigma_c$ and the zero temperature limiting behavior 
behavior for $\sigma>\sigma_c$ by generalizing Eq. (\ref{eq:sigmavsgammadot})
to a form similar to equation (\ref{chattoraj}), namely
\begin{equation}
    \sigma(\dot{\gamma},T) = \sigma(\dot{\gamma},T=0) - \left [\frac {T}{C}\ln\left (\frac{C'T^\psi}{\gdot}\right)\right]^{1/\alpha}.
\label{interpolacion}
\end{equation}
where $\sigma(\dot{\gamma},T=0)$ is expected to behave as
\begin{equation}
\sigma(\dot{\gamma},T=0)=\sigma_c + C_0\dot{\gamma}^{1/\beta}
\label{sigma_de_T0}
\end{equation}
In addition to reducing to the standard flowcurve
at $T=0$ and to the exponential activation formula
when $T\gg \gdot ^{1/\psi}$, Eqs. (\ref{interpolacion}), (\ref{sigma_de_T0}) 
are fully compatible with the general thermally 
activated behavior (Eq. (\ref{eq:yielding_thermalrounding})).

\begin{figure}[t!]
\includegraphics[width=\columnwidth]{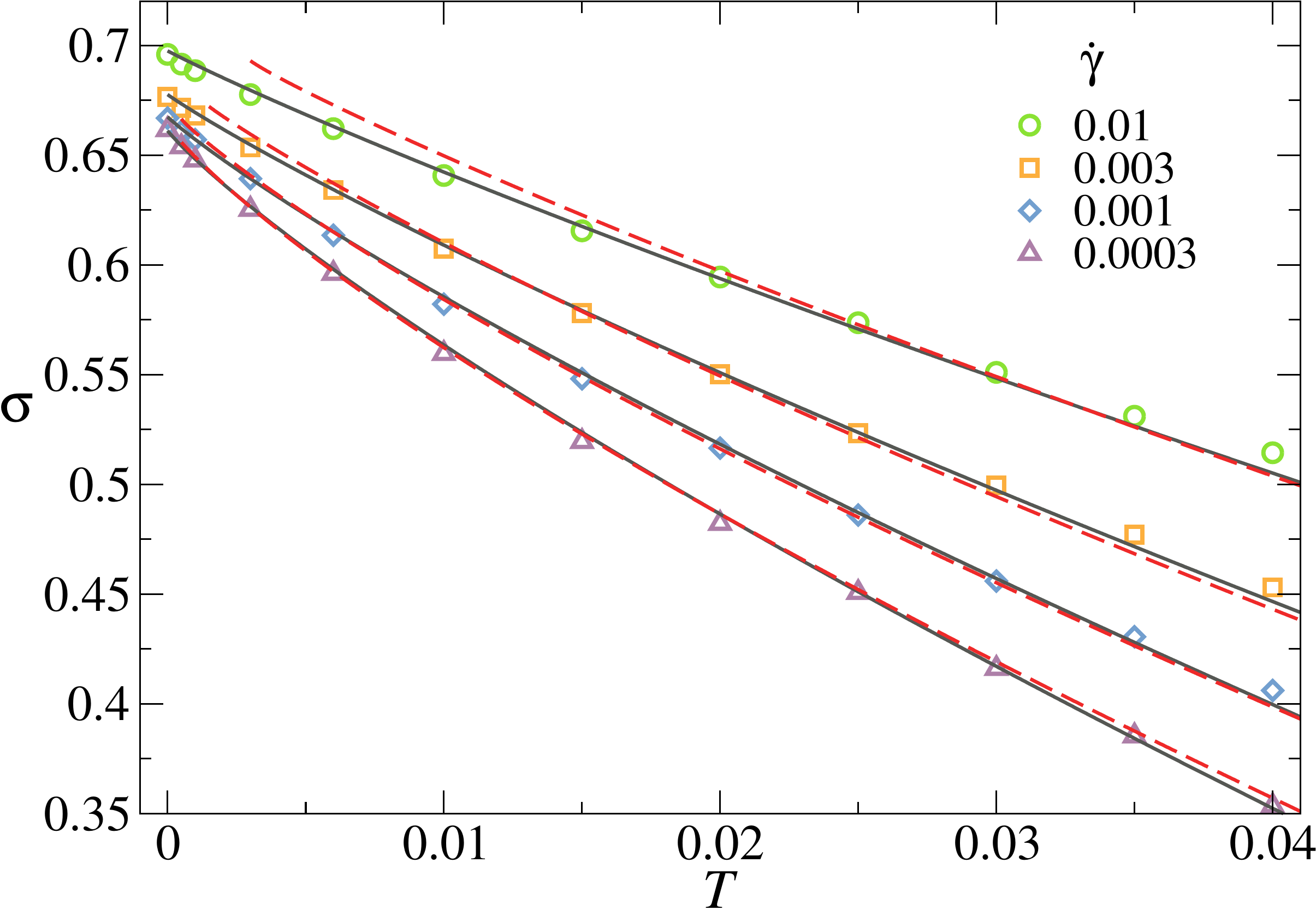}
\caption{
Simulation results for $\sigma$ as a function of $T$ at different 
$\gdot$ levels for the data corresponding to Fig.\ref{fig:256smooth}.
The red-dashed lines correspond to the prediction of Eqs.~(\ref{interpolacion}), (\ref{sigma_de_T0})
with $\alpha=3/2$, $\psi=4/3$, $\sigma_c=0.6535$, $C_0=0.44$, $C=1.70$, $C'=27.47$
(notice that this prediction cannot be extended at temperatures lower than $\sim (\gdot/C')^{1/\psi}$).
The gray full lines correspond to the expression (\ref{interpolacion_full})
with $\alpha=3/2$, $\psi=4/3$, $\sigma_c=0.6535$, $C=0.45$, $C'=1.41$, $\kappa=0.375$.
}
\label{JS_form}
\end{figure}
It is interesting to check our numerical data against
expression (\ref{interpolacion}).
For the sake of concreteness we only show the results for the Hamiltonian 
model in the case of smooth potentials.
Using the same parameters that were used to construct Fig.~\ref{fig:256smooth}, 
we obtain the red-dashed curves shown in Fig.~\ref{JS_form} 
(notice the analogy with Fig.~2 in \cite{JohnsonPRL2005}).
We see that the fitting to the numerical values provided by expression 
(\ref{interpolacion}) (adjusting constants $C_0$, $C$ and $C'$ and using the 
appropriate exponents for this case, namely $\alpha=3/2$ and $\beta=2$)
is in fact very good if the temperature is not too small.
However in this limit Eq. (\ref{interpolacion}) cannot be correct as the {\em {log}} becomes negative.
A full range approximate interpolation scheme can be easily obtained by transforming 
(\ref{interpolacion}) to 

\begin{equation}
    \sigma(\dot{\gamma},T) = \sigma(\dot{\gamma},T=0) - \left [\frac {T}{C}\ln
    \left (C'\left (\frac{T^\psi}{\gdot}\right )^{\kappa}+1\right)\right]^{1/\alpha}.
\label{interpolacion_full}
\end{equation}
This regularization of the {\em log} is similar to the one that is known to work very well in the 
Prandtl-Tomlinson model \cite{MuserPRB2011},
and provides a much better fitting at low $T$ to the data in Fig. \ref{JS_form}, as indicated by the gray full lines.

\begin{figure}[t!]
\center
\includegraphics[width=0.95\columnwidth]{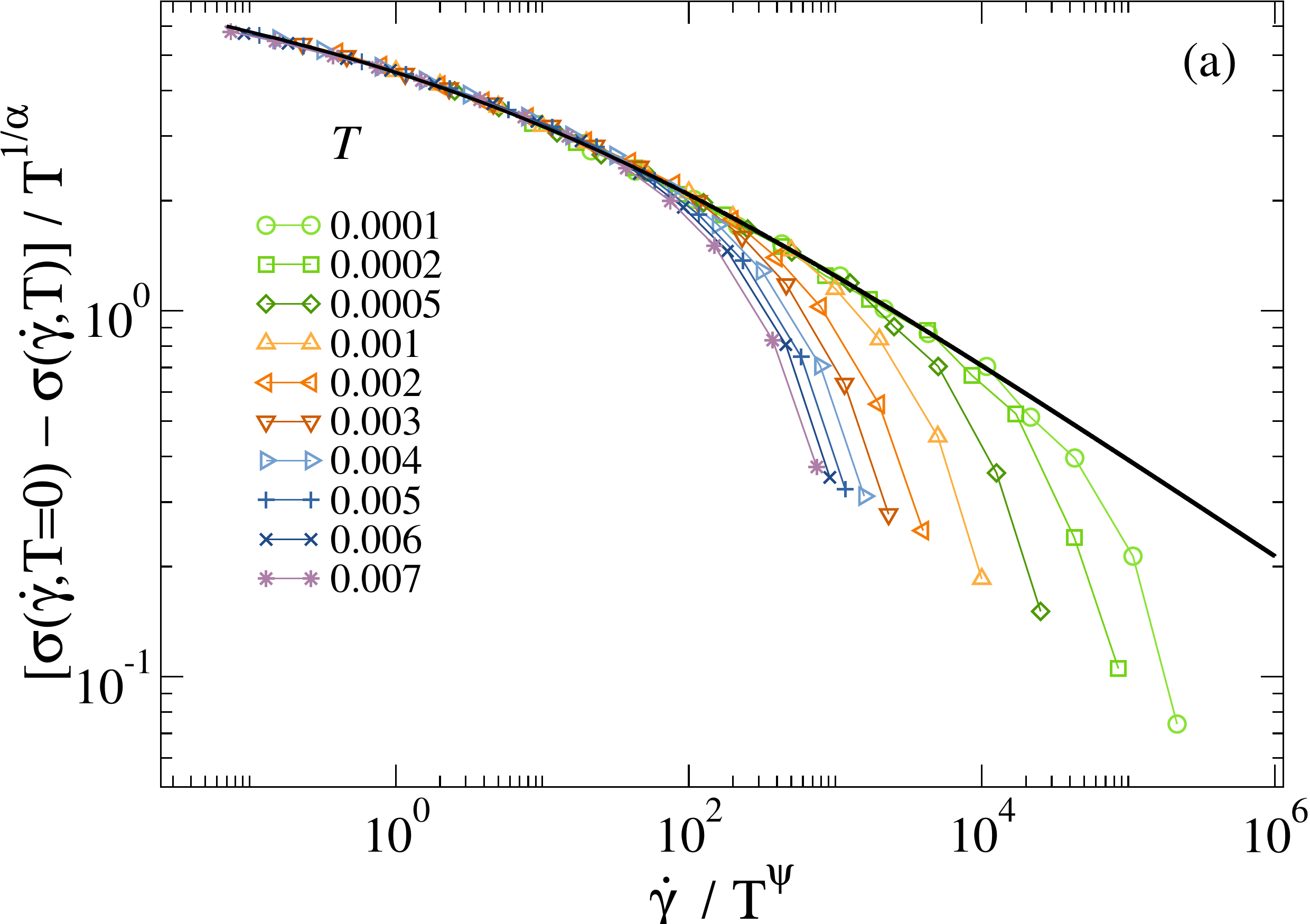}
\includegraphics[width=0.95\columnwidth]{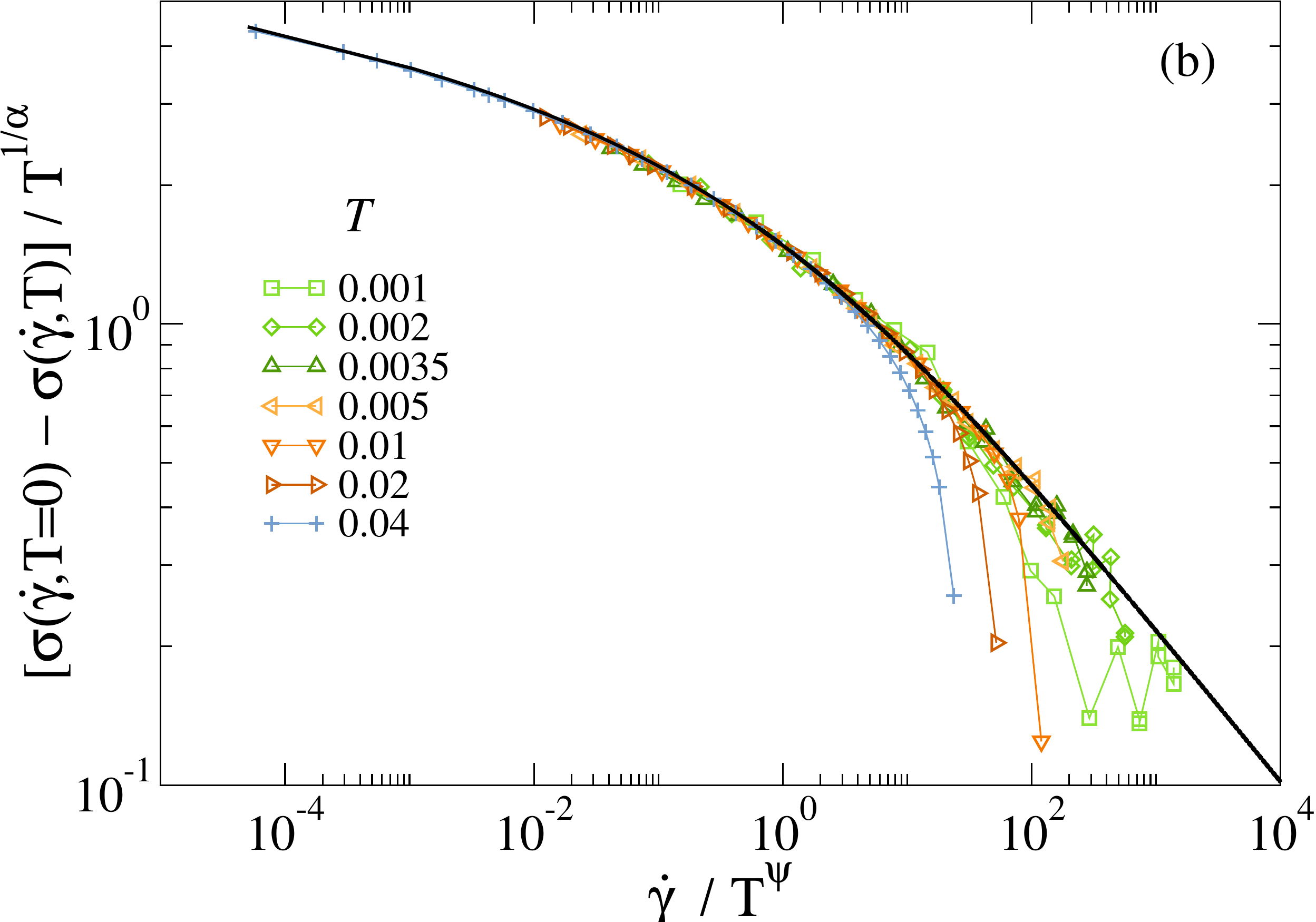}
\includegraphics[width=0.95\columnwidth]{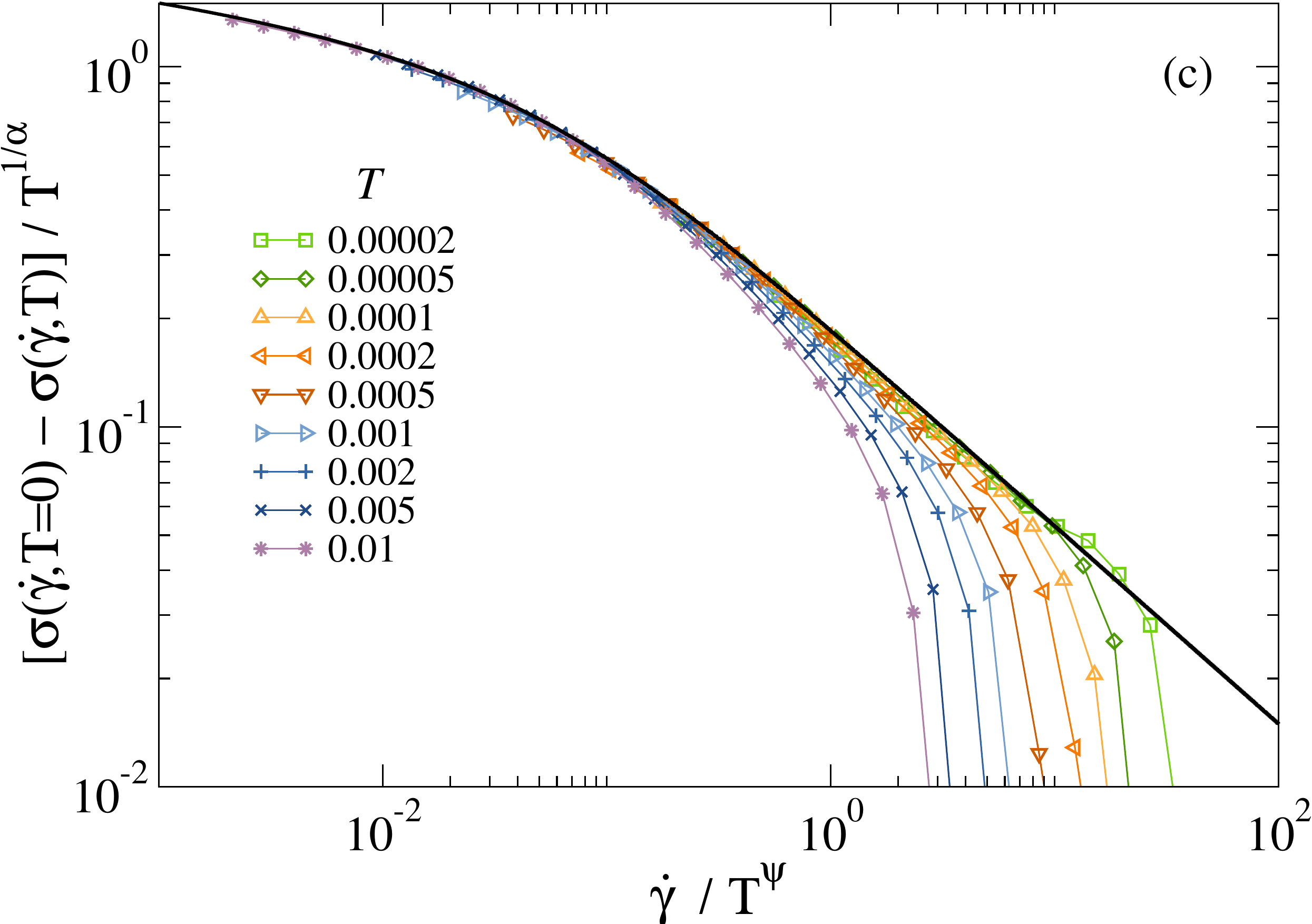}
\caption{
Plot of $[\sigma(\gdot,T=0)-\sigma(\gdot,T)]/T^{1/\alpha}$ vs $\gdot/T^{\psi}$, for 
(a) the one particle model and 
(b) the extended Hamiltonian model (system size 256 $\times$ 256), 
with smooth potentials ($\psi=4/3$), and for 
(c) the EP model with instantaneous events (system size 2048 $\times$ 2048), 
for which $\psi=3/4$.
Different temperatures are included and fall into the same mastercurve.
In all cases the points are fitted by an expression of the form
$y=\left(C^{-1} \log(C'/x^{\kappa}+1)\right )^{1/\alpha}$.
(one particle: $\alpha=3/2$, $\kappa =0.4$, $C=0.18$, $C'=4.5$; 
Hamiltonian: $\alpha=3/2$, $\kappa =0.5$, $C=0.61$, $C'=2$;
EPM: $\alpha=2$, $\kappa =1.1$, $C=2.25$, $C'=0.08$) 
}
\label{fig:log23}
\end{figure}
The interpolation scheme of Eq. (\ref{interpolacion_full}) suggests to plot the values 
of $y\equiv \left[\sigma(\gdot, T=0)-\sigma(\gdot,T)\right ]/T^{1/\alpha}$ vs. 
$x\equiv \gdot/T^\psi$ for the one-particle and the full extended model to compare in detail 
the effect of temperature in  both cases. 
In fact, this plot is of the form
\begin{equation}
    y =  \left [\frac 1 {C}\ln
    \left (\frac {C'}{x^{\kappa}}  +1\right)\right]^{1/\alpha}
\label{interpolacion_adimensional}
\end{equation}
and must lie on a single master curve if the thermal rounding scaling is satisfied.
Results are presented in Fig.~\ref{fig:log23} for Prandtl-Tomlinson, Hamiltonian and 
elastoplastic models. 
We see that the data remarkably collapse on a single master curve as temperature is reduced~\footnote{In order to be 
able to subtract curves at fix $\gdot$, the values $\sigma(\gdot,T=0)$ are taken from 
the analytical fit of the sparse data obtained at arbitrary fix stress values. 
This explains the deviation from the mastercurve of the points of larger temperatures 
when $[\sigma(\gdot,T=0)-\sigma(\gdot,T)]/T^{1/\alpha}$ approaches zero.}, 
and also that in the three cases this curve  is accurately fitted by and expression of 
the form (\ref{interpolacion_adimensional}).
The {\em log-two-thirds} behavior (for $\alpha=3/2$) of this expression 
is the one expected based on the 
interpolation formula Eq.~(\ref{interpolacion}). 
The power of $x$ and the $+1$ added term inside the {\it log} give a much better 
crossover to the power law decay as $x\to\infty$. 
In any case, the remarkable result is that the same kind of analytical expression provides a 
very good fitting of the results for the one particle model and for the full 
extended model.

\section{Summary and Discussion}
\label{sec:discussion}

We have addressed the problem of the thermal rounding of the yielding 
transition of amorphous materials in a comprehensive theoretical framework, 
both including different modeling approaches and analytical arguments and 
targeting the interpretation of important phenomenological laws based on 
experimental data from a new perspective. 
In particular, we have considered two different numerical approaches 
consisting in spatially extended models that describe stress and strain 
in the system at a coarse-grain level, where the elastic interaction at 
finite distance is incorporated through the use of the Eshelby quadrupolar kernel.
In one case, coined ``Hamiltonian'', the full dynamics has the 
form of an overdamped equation of motion for the local strain.
In this case temperature is included in a standard way through the 
addition of a Langevin stochastic term to the equations of motion.
The second case corresponds to the purely phenomenological approach of
elasto-plastic models, where the elastoplastic blocks of the system switches 
between solid and fluidized states according to local rules that take into 
account their mechanical stability.
In this case we include temperature as an Arrhenius-like activation allowing 
one block to fluidize even when its stress is lower than the local yielding threshold.

Our first main result has been to extend the compatibility between these
modeling approaches to the case of finite temperatures.
Previous works~\cite{FernandezAguirrePRE2018, FerreroSM2019} indicated 
that at $T=0$ both approaches display the same qualitative behavior 
at both quasistatic and finite strain rate deformations.
In particular, at $T=0$, irrespective of the model particularities and 
close to the critical stress $\sigma_c$, all flowcurves group in only one 
of two families~\cite{FernandezAguirrePRE2018,FerreroSM2019}: 
the one corresponding to ``cuspy'' disordered potentials 
(\emph{equiv.} uniform yielding rates in EPMs) with $\beta=3/2$
or the one corresponding to ``smooth'' disordered potentials 
(\emph{equiv.} progressive yielding rates in EPMs)  with $\beta=2$.
For finite temperatures $\gdot$ is different from zero even below $\sigma_c$ 
displaying an exponential activation of the form 
\begin{equation}
\gdot \sim \exp [-C(\sigma_c-\sigma)^\alpha/T]~~~~~~~~(\sigma<\sigma_c)
\label{exponencial}
\end{equation}
The value of $\alpha$ encodes details of the quenched stochastic potential in the 
Hamiltonian case, or the activation rates as a function of stress in the EPMs. 
The expected $\alpha$ value corresponding to the most realistic case of a smooth quenched 
potential (in Hamiltonian modeling) is $\alpha=3/2$, and is the one that should be expected
in molecular dynamics simulations and experiments.

The second main result of our work is the observation that the flowcurves, 
in a finite interval around $\sigma_c$, and for finite temperatures 
(at least when $T$ is not ``too large'') are very well described by the 
thermal rounding scaling of Eq. (\ref{eq:yielding_thermalrounding}).
This scaling extends the exponentially activated regime for 
$\sigma<\sigma_c$ to a full interval around $\sigma_c$. 
In fact we conclude that the Johnson and Samwer relation, Eq. (\ref{eq:JandS}),
can be derived from Eq. (\ref{eq:yielding_thermalrounding}) and recovered
with all the numerical approaches we have implemented.
The connection of the thermal rounding scaling proved in this work with the 
phenomenological results gathered in~\cite{JohnsonPRL2005} strongly suggests 
that our conclusions, besides a pure theoretical interest, are relevant 
to the study of thermal rounding of real amorphous materials where a flowcurve 
can be actually measured, for example, in colloidal glasses~\cite{BonnRMP2017,PetekidisJPCM2004}.

The thermal rounding scaling of eq.\ref{eq:yielding_thermalrounding} is predicted from 
models that assume a well defined temperature $T$ entering either through a Langevin 
noise or through Arrhenius activation at the corresponding coarse grained level for each case. In the case of amorphous materials with mesoscopic constituents (colloidal glasses/gels,
emulsions, foams), whether such $T$ should correspond to the bath temperature or a
thermodynamically well defined effective temperature of the material~\cite{BerthierJCP2002,Cugliandolo1997} 
(that incorporate non-equilibrium fluctuations) remains an open and interesting question. 
Nonetheless, the agreement of our model predictions with the Jhonson \& Samwer
phenomenological scaling strongly suggests that, at least for metallic glasses, 
the putative effective temperature $T$ must be equal or proportional to the 
experimentally measured temperature. 
In addition, any effective temperature playing the role of $T$ in
Eq.~\ref{eq:yielding_thermalrounding} should have a negligible $\dot\gamma$ dependence,
otherwise the whole scaling would fail.

We have also shown that the thermal rounding scaling is analytically satisfied 
in the case of a single particle driven on a disorder quenched potential, 
under the action of a mechanical and a thermal noise. 
This concomitance between the thermal rounding behavior of the one-particle model 
and that of the full extended model, which extends also to the detail of the analytical 
form of the full $\sigma(\gdot,T)$ curve, reinforces our view that the spatially 
distributed simulation outcomes admit a very accurate description in terms of a 
one-particle system. 
In the end, this is an additional indication that the yielding transition of amorphous 
materials in finite dimensions, at least up to the point in which it is captured by the 
present kind of models, can be described effectively as a mean field transition.  

\begin{acknowledgments}
EEF acknowledges support from PICT-2017-1202 and ABK from PICT-2016-0069. 
We also acknowledge support from UNCuyo-2019 06/C578.
\end{acknowledgments}

\appendix

\section{Model description}
\label{app:modeldescription}

\subsection{Hamiltonian Model}

The Hamiltonian description of the yielding transition and plastic behavior has 
been already presented in~\cite{JaglaJSTAT2010,JaglaPRE2007,FernandezAguirrePRE2018}.
We provide here a short description for completeness.
It considers the symmetric, linearized elastic strain tensor of 
the material $\varepsilon_{ij}(r)$ at different positions $r$ in 
the sample.
It assumes a relaxational dynamics that tends to minimize the free 
energy of the system, in the form
\begin{equation}
\eta \frac{\partial \varepsilon_{ij}}{\partial t}= {-}\frac{\delta F(r)}{\delta \varepsilon_{ij}} +\Lambda_{ij}\varepsilon_{ij}+\sigma_{ij} .
\label{tensorial}
\end{equation}
Here, $F\equiv \int d^d r f(\varepsilon_{ij})$ is the total free energy, 
obtained by spatial integration of a free energy density $f$. 
Note that $f$ is {\em local} in $\varepsilon_{ij}$.
$\Lambda_{ij}$ are Lagrange multipliers that are necessary to fulfill 
internal constrains among the $\varepsilon_{ij}$, usually referred to as Saint-Venant
compatibility conditions~\cite{JaglaJSTAT2010,JaglaPRE2007}. 
The $\sigma_{ij}$ are externally applied stresses with different symmetries. 
In the form given by \ref{tensorial}, this is already a model that can be applied 
to concrete calculations in a fully tensorial framework, once the form of 
$f(\varepsilon_{ij})$ is defined~\cite{JaglaPRE2020}.
However, in the case in which the externally applied stress is homogeneous and 
of definite symmetry, a further {\em approximate} transformation can be proposed, 
as follows. 
If, for simplicity, we call $\sigma$ the applied external stress, and $e$ the 
corresponding component of the strain field, we can (under certain conditions~\cite{JaglaPRE2020}) integrate out the remaining components of the $\varepsilon_{ij}$ tensor, and 
arrive at a scalar model  for $e(r)$. 
Switching now to a notation in which the latin indexes label spatial positions 
in the sample, this scalar model reads (we take $\eta=1$ from now on)
\begin{equation}
\frac{\partial e_i}{\partial t}={-}\frac{\delta F(e)}{\delta e_i} +\sum_{j}G_{ij}e_{j}+\sigma.
\label{model}
\end{equation}
Note that the original compatibility conditions have transformed in the {\em non-local} interaction term mediated by the kernel $G_{ij}$.
The detailed derivation shows that $G_{ij}$ is nothing but the Eshelby interaction also used in EPM's (see next Section).
All that remains to define our model is to specify the form of the free energy $F$. First of all, notice that $F$ is a sum of local term over different parts of the sample, i.e, 
\begin{equation}
F(e)=\sum_i V_i(e_i)
\end{equation}
Since we are interested in modeling an amorphous, disordered material $V_i(e_i)$ will be chosen in such a way that it  describes the local thresholding behavior
of a small piece of the amorphous material under deformation.
The functions $V_i$ have minima at different values of $e$ representing local
equilibrium states. The functions $V_i$ are stocastically defined, in an uncorrelated manner for each site $i$. 

\begin{equation}
\frac{de_{\bf q}}{dt}=-\sum_i V'_i(e_i)|_{\bf q}+G_{\bf q}e_{\bf q}
\label{modeloq}
\end{equation}
$G_{{\bf q}=0}$ is taken as zero in a stress conserved dynamics.
The uniform mode in Eq.(\ref{model}) is thus directly found from
\begin{equation}
\dot\gamma\equiv  \frac{d\overline{e_i}}{dt}=\overline{-V'_i(e_i)}+\sigma
\label{gammadot}
\end{equation} 
that defines the global strain rate $\dot\gamma$.
Finally, the last remaining point is related to the incorporation of temperature. In the present model, there is a simple and natural way to incorporate temperature, namely
in the form of a stochastic (Langevin) force, added to the right of Eq. \ref{model}, that finally reads
\begin{equation}
\frac{\partial e_i}{\partial t}=- \frac{d V_i}{d e_i} +\sum_{j}G_{ij}e_j+\sigma+\sqrt {T} \xi_i(t)
\label{model_mas_t}
\end{equation}
with the stochastic term $\xi(t)$ satisfying
\begin{eqnarray}
\langle \xi_i(t) \rangle&=&0\\
\langle \xi_i(t)\xi_j(t') \rangle&=&2\delta(t-t')\delta_{ij}
\end{eqnarray}

We numerically simulate Eq. \ref{model_mas_t} 
for two particular on-site periodic potentials referred to as 
``cuspy''and ``smooth'' potentials. 
They are constructed as a concatenation of parabolic (cuspy) or sinusoidal (smooth)
pieces, in consecutive intervals of the $e$ axis. Each interval is characterized by its left and right border, $e_l$, $e_r$, in such a way that the force derived from the potential in a particular interval is given in terms of  $e_0\equiv (e_l+e_r)/2$ and $\Delta\equiv (e_r-e_l)$ by
\begin{equation}
  -\frac{dV}{de}=\begin{cases}
    e_0-e, & \text{``cuspy potential''},\\
    \frac{\Delta}{2\pi}\sin(2 \pi (e_0-e)/\Delta), & \text{``smooth potential''},
  \end{cases}
  \label{eq:potentials2d}
\end{equation}
The value of $\Delta$ for each interval is taken from a flat distribution within the interval $[1,2]$. 
It is clear from its definition, that the cuspy (smooth) potential has a discontinuous (continuous)
force  between consecutive intervals of definition.

We integrate the equation of motion using a first order Euler method with a temporal time 
step $\delta t=0.1$. 
All results presented correspond to square samples with periodic 
boundary conditions. The flowcurve is determined starting from the largest values of $\sigma$, and progressively reducing it, while the strain rate 
is calculated from Eq. \ref{gammadot}. In this way we get rid of issues associated to sample preparation that would appear if the smallest $\sigma$ values were simulated first.

\subsection{Elasto-plastic model}

When referring to elasto-plastic models (EPMs), we consider amorphous materials at 
a coarse-grained-level description, laying in between the particle-based simulations 
and the continuum-level description. 
Full background, context and historical development of EPMs can be found in~\cite{NicolasRMP2018}.
The amorphous solid is represented by a coarse-grained scalar stress field $\sigma(\br,t)$, at spatial position $\br$ and time $t$ under an externally applied shear strain. Space is discretized in blocks. At a given time, each block can be ``inactive'' or ``active'' (i.e., yielding). This state is defined by the value of an additional variable: $n(\br,t)=0$ (inactive), or $n(\br,t)=1$ (active).

We define our EPM in $2$-dimensions discretized on a square lattice, 
and each block $\sigma_i$ subject to the following evolution in real space
\begin{equation}\label{app_eq:eqofmotion1}
\frac{\partial \sigma_i(t)}{\partial t} =
  \mu\dot{\gamma}^{\tt ext}  +\sum_{j} G_{ij} n_j(t)\frac{\sigma_j(t)}{\tau} ;
\end{equation}
where $\dot{\gamma}^{\tt ext}$ is the externally applied strain rate,  
and the kernel $G_{ij}$ is the Eshelby stress propagator~\cite{Picard2004}.

It is convenient to explicitly separate the $i=j$ term in the previous sum, as
\begin{equation}\label{app_eq:eqofmotion2}
\frac{\partial \sigma_i(t)}{\partial t} =
  \mu\dot{\gamma}^{\tt ext}  - g_0 n_i(t)\frac{\sigma_i(t)}{\tau} + \sum_{j\neq i} G_{ij} n_j(t)\frac{\sigma_j(t)}{\tau} ;
\end{equation}
where $g_0\equiv -G_{ii} > 0$ (no sum) sets the local stress dissipation rate for an 
active site.
The form of $G$ is $G(\br,\br') \equiv G(r,\varphi)\sim\frac{1}{\pi r^2}\cos(4\varphi)$
in polar coordinates, 
where $\varphi \equiv \arccos((\br-\br')\cdot\br_{\dot{\gamma}^{\tt (ext)}})$ and
$r \equiv \left|\br-\br'\right|$. 
For our simulations we obtain $G_{ij}$ from the values of the propagator in Fourier 
space $G_{\bf q}$, defined as
\begin{equation}
G_{\bf q} = -\frac{4q_x^2q_y^2}{(q_x^2+q_y^2)^2}.
\label{app_eshelby_kernel}
\end{equation}
for $\bf q\ne 0$ and 
\begin{equation}
G_{\bf q=0}=-\kappa
\label{app_eshelby_kernel_q0}
\end{equation}
with $\kappa$ a numerical constant set to 1.

The elastic (e.g. shear) modulus $\mu=1$ defines the stress unit, and the mechanical
relaxation time $\tau=1$, the time unit of the problem.
The last term of (\ref{app_eq:eqofmotion2}) constitutes a \textit{mechanical noise}
acting on $\sigma_i$ due to the instantaneous integrated plastic activity
over all other blocks ($j\neq i$) in the system.
The picture is completed by a dynamical law for the local state variable $n_i=\{0,1\}$. 
Here is where the thermal activation for $T>0$ steps in.
In the athermal case, when the local stress overcomes a local yield stress, 
a \textit{plastic event} occurs (the block becomes ``active'') with a given 
probability, usually constant (see \cite{FerreroSM2019} for different alternatives).
But when $T>0$ we also expect activation to occur with a finite probability
even when $\sigma_i<\sigmaY_i$.
The block ceases to be active when a prescribed criterion is met. 
When the plastic event has a finite duration, a local memory is coded in the 
system configuration, defining a dynamics that is typically non-Markovian.
In this work we have used the following rules for sites activation and deactivation:

\begin{equation}\label{app_eq:activationrulesEPM}
n_i : \begin{cases} 
0 \rightarrow 1 & \mbox{instantaneously~} \mbox{\quad if \quad} \sigma_i \geq \sigmaY \\
0 \rightarrow 1 & \mbox{with probability per unit time~} \\
& \exp(-(\sigmaY_i-\sigma_i)^\alpha/T) \mbox{\quad if \quad} \sigma_i <\sigmaY \\
0 \leftarrow 1  & \mbox{at a rate~} \tau_{\tt off}^{-1} 
\end{cases}
\end{equation}
\noindent where $\alpha$ and $\tau_{\tt off}$ are parameters 
and the $\sigmaY_i$ variables are randomly sorted after each
local yield event to be $1 + 0.1 r_{\tt exp}$, with $r_{\tt exp}$ a random 
number taken from an exponential distribution of average unity.
The case of instantaneous stress release corresponds to $\tau_{\tt off} \to \infty$,
otherwise we have set  $\tau_{\tt off}=1$.
As discussed in~\cite{FerreroSM2019}, the case of EPMs with \emph{uniform}
local yield rates (i.e., constant, as in this case) can be directly related 
to the case of cuspy potentials in the Hamiltonian model.
The $\beta$ exponent of the athermal flowcurve results identical in both approaches.
We then believe that the choice of the parameter $\alpha$ in the thermal
activation rule is not arbitrary but should respect the same analogy among
model approaches.
Therefore, here we use $\alpha=2$ which is the barrier exponent in a parabolic 
potential.
On the other hand, the case of smooth potentials in the Hamiltionian approach 
is analogous to the case of \emph{progressive} local yield rates~\cite{FerreroSM2019}
in EPMs.
In that case the block activation is stochastic by definition. 
We have avoided here to combine the stochasticity of both progressive rates 
(e.g. $\tau_{\tt on} \sim (\sigma_i-\sigma_i^{\tt y})^{-1/2}$) and thermal 
activation, and choose to show only the uniform rate case for simplicity.
But such a combination is possible to do and in that case we would use $\alpha=3/2$
as the barrier exponent for the thermal activation in~\eqref{app_eq:activationrulesEPM}.

\section{Scaling for a single particle in a potential}\label{app:pt}
\label{sec:appendixoneparticle}

\begin{figure}[t!]
\includegraphics[width=6cm,clip=true]{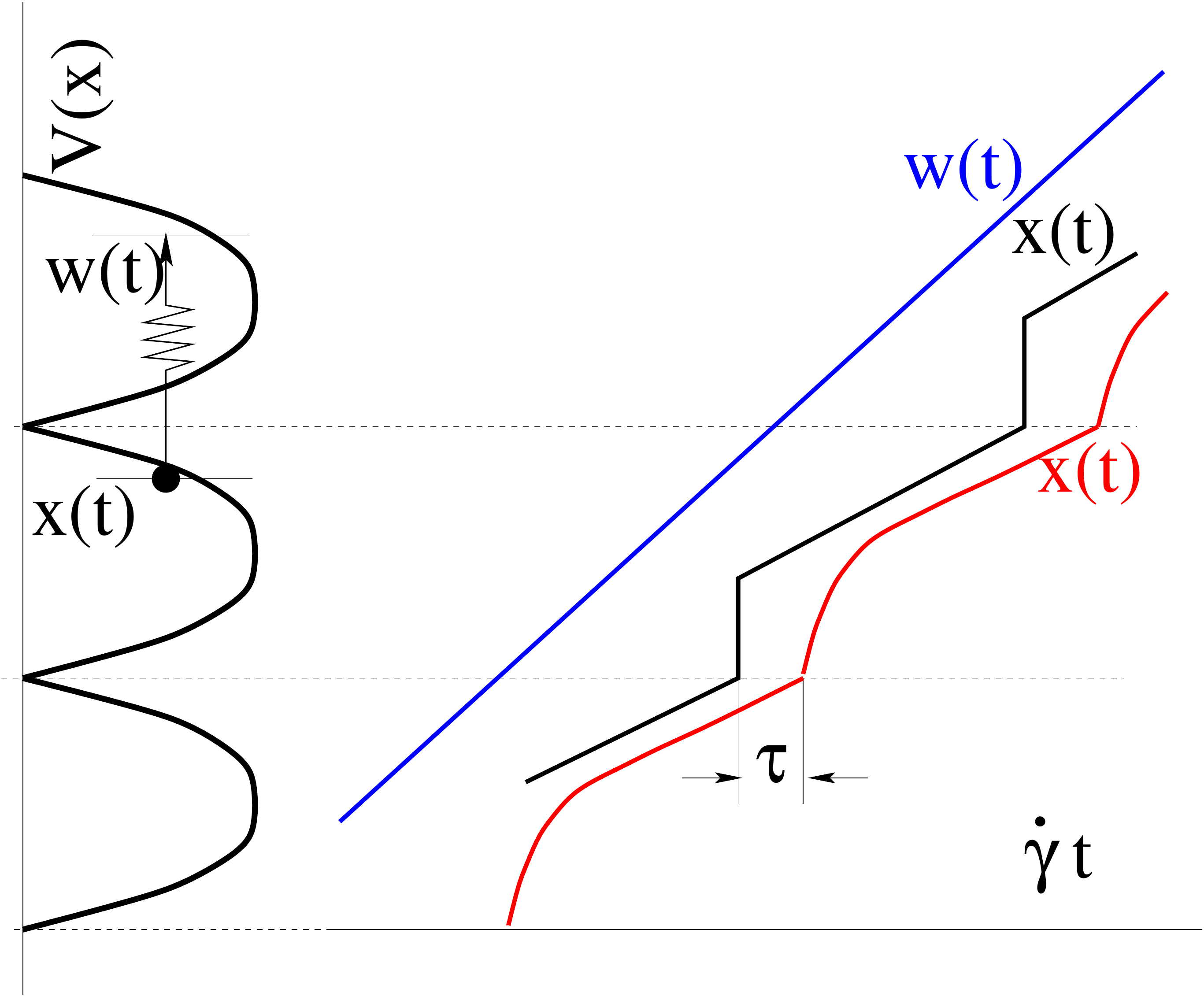}
\caption{Schematic representation of the dynamics of the Prandtl-Tomlinson model at $T=0$ and $a=0$.
On the left, a periodic potential $V(x)$ constructed by a concatenation of 
parabolas, with a particle moving on it, is depicted with a 90º rotation 
for visualization.
On the right, the driving $w(t)$ and particle
position $x(t)$ are shown for $\gdot\to 0$ (black) and $\gdot> 0$ (red).
The value of $\sigma$ is obtained as the average of $k_0(w(t)-x(t))$. This value is larger
when $\gdot>0$ than when $\gdot\to 0$.
}
\label{fig:pt_tau}
\end{figure}

In this appendix we derive the scaling form of the flowcurve for a 
single particle stochastically driven in a potential in the presence
of thermal noise.  
The stochastic driving is composed of a fix strain-rate, or velocity,
plus a mechanical noise characterized by a Hurst exponent $H$.
The derivation generalizes the one of Ref.~\cite{jagla2018prandtl} to 
the case of finite temperatures.
The $x$ variable of the system (particle position) follows Eqs.\eq{eq_1pla1} 
and \eq{eq_1pla2}, that we repeat here for convenience
\begin{eqnarray}
\frac{dx}{dt}&=&-\frac{dV}{dx}+k_0\left(w(t)-x\right) +\sqrt{T}\eta_0(t) \label{app1}\\
\frac{dw}{dt}&=&b\dot \gamma+a{\dot\gamma}^H \eta_H(t)\label{app2}
\end{eqnarray}
The system is driven imposing a constant value of $\dot\gamma$.
At a given time $t$, $\eta_H$ is a random variable sorted as described 
in Eq.~\ref{eq:etaH}.
The stress $\sigma$ at any moment is defined as 
$\sigma\equiv k_0(\overline{w(t)-x(t)})$, where $k_0$ is a model parameter.

If $T=0$ and $k_0<\max_x (-V''(x))$ there is a finite critical 
stress value $\sigma_c$ when $\dot\gamma$ tends to zero. 
In a general case, for small values of $T$ and $\dot\gamma$, 
$\sigma$ will be close to $\sigma_c$ (i.e., $|\sigma-\sigma_c|/\sigma_c\ll 1$).
We want to find a scaling relation between $T$, $\dot\gamma$, and $\sigma-\sigma_c$ 
close to the critical point in which all these three variables are vanishing.
The idea of the calculation is as follows.
Consider the evolution of the variable $x$ as a function of $\dot\gamma t$ 
at $T=0$, $a=0$, and for a vanishingly small $\dot\gamma$, as depicted in Fig.~\ref{fig:pt_tau}. 
As the particle advances in the potential $V(x)$, ``jumps'' in $x$ occur at the 
transition points between local basins (black line for $x(t)$ in Fig.~\ref{fig:pt_tau}).
The value of the stress is proportional to the average of $w(t)-x(t)$, and in the case 
of $T=0$ and vanishing $\dot\gamma$ will be $\sigma_c$. 

\begin{figure}[t!]
\includegraphics[width=\columnwidth]{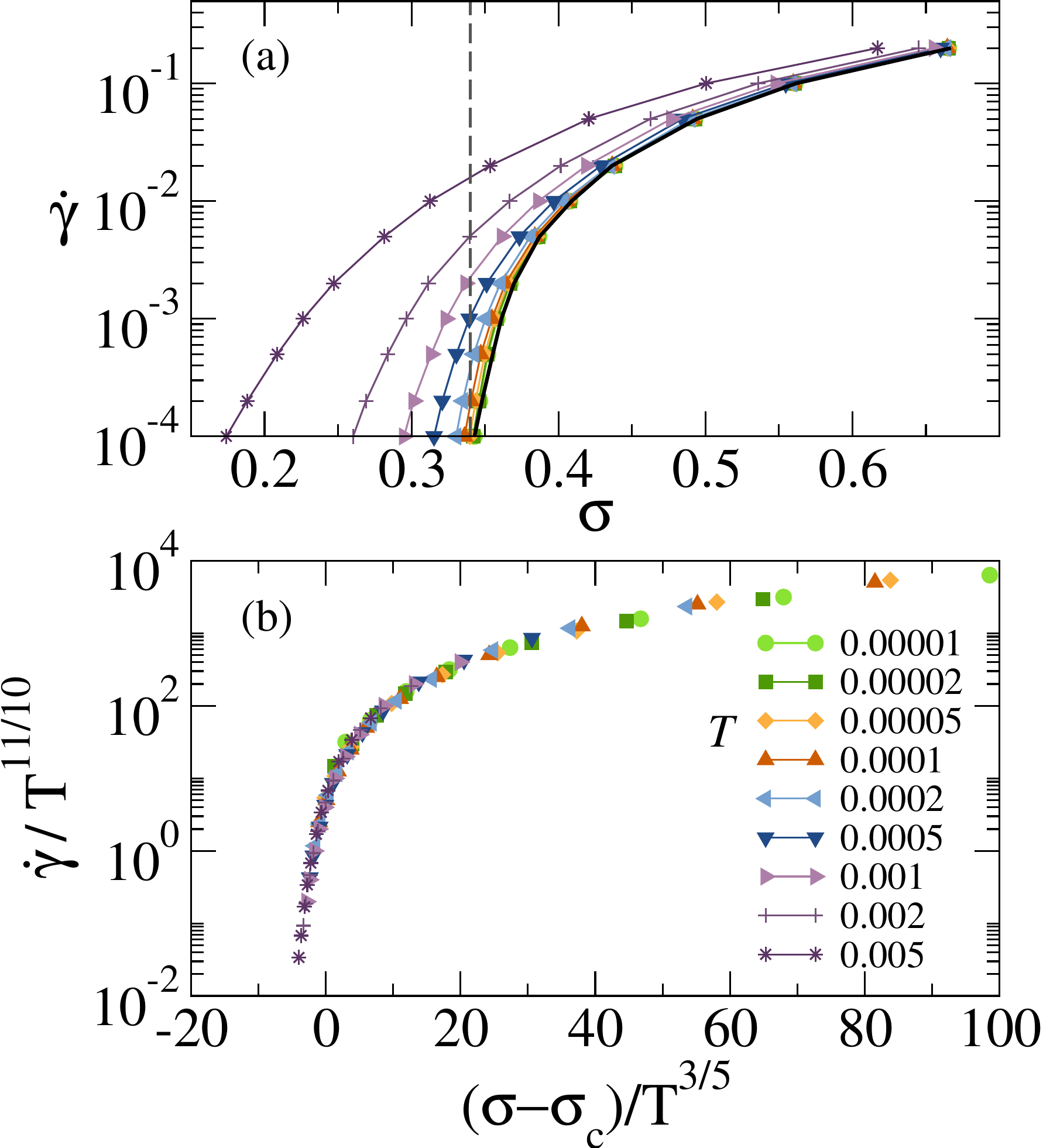}
\caption{
(a) Flow-stress curves for one particle in a periodic potential, with mechanical noise 
of $H=2/3$ at different temperatures. 
(b) Master curve 
(Eq.\ref{eq:yielding_thermalrounding})
using the exponents corresponding to to the $\omega=3/2$ case, $\psi=11/10$ 
and $1/\alpha=3/5$ from Eq.(\ref{eq:exponentsPT}), and $\sigma_c=0.34$.
}
\label{fig:1pla1intermedia}
\end{figure}

For finite but small $T$ and $\dot\gamma$, the evolution of $x$ will be 
close but not exactly equal to the previous case. 
The average of $w(t)-x(t)$ will be different, in particular due to the finite 
$\dot\gamma$, but also due to a finite temperature.
The main effect on $\sigma$ can be understood due to a shift in the transition point
from one basin to the next one.
Now, it is not necessarily true that $x$ will jump exactly when reaching the cusp 
edge (or the maximum derivative for a smooth potential).
We encode this time shift in a variable $\tau$ (see Fig.~\ref{fig:pt_tau}).
The change in stress $\delta \sigma\equiv \sigma-\sigma_c$ can be estimated 
as the fraction of time that $\tau$ represents of the total time needed 
to traverse a basin (a period of the potential).
Being the latter $(b\dot\gamma)^{-1}$, we find 
$\delta\sigma\sim \tau b\dot\gamma$. 
The following step to quantify the change in $\sigma$ is to obtain the scaling 
behavior of $\tau$ from Eqs. \ref{app1} and \ref{app2}. 
Taking into account the importance of the transition points, we first rewrite 
Eq. \ref{app1} close to these points using $-dV(x)/dx\simeq Ax^{\omega}$ as
\begin{equation}
\frac{dx}{dt}=A|x|^\omega+ k_0 w(t) +\sqrt{T}\eta_0(t)
\label{app3}
\end{equation}
Note that $\omega=1$ describes the situation of a potential formed by consecutive 
parabolic pieces, while $\omega=2$ corresponds to smooth potential. 
The value of $\tau$ that we search for must be expressible in term of 
the parameters appearing in Eqs. (\ref{app2}), (\ref{app3}). 
These parameters are $A$, $a$, $b$, $\dot\gamma$ and $T$. 
From these four parameters, three (and non-redundantly only three) quantities 
with time dimensions can be constructed. 
They can be taken to be:
{\large
\begin{eqnarray}
t_1 & \equiv &   a^{\frac{1-\omega}{\omega(1-H)}}  b^{\frac{(\omega-1)H}{\omega(1-H)}} A^{-\frac{1}{\omega}} \\
t_2& \equiv&   a ^ {\frac{2-3\omega}{\omega(1-H)}} b^{\frac{2H\omega-2H+\omega}{\omega(1-H)}} A^{-\frac{2}{\omega}}\dot\gamma\\
t_3& \equiv&   T^{\frac13}b^{\frac23}\dot\gamma^{-\frac23}
\end{eqnarray}
}
On dimensional grounds, the value of $\tau$ can be expressed in general in the form 
\begin{equation}
\tau=t_1F(t_2/t_1,t_3/t_1)
\end{equation}
where $F$ is an unknown function. 
From here we can write 
\begin{equation}
\delta\sigma=\dot\gamma b t_1F(t_2/t_1,t_3/t_1)
\label{ttt}
\end{equation}
One more condition can be used to specify this expression. 
For small values of $\dot\gamma$, the $\sim\dot\gamma^H$ in Eq.~(\ref{app2}) 
must dominate over the $\sim\dot\gamma$ term. 
In other words, this means that in the final expression for $\delta\sigma$ the 
dependence on $b$ has to drop out, it has no relevance. 
This allows to eliminate one of the dimensionless variables in Eq.~(\ref{ttt}). 
After some algebra we can finally write the dependence of $\delta\sigma$ on 
$\dot\gamma$ and $T$ as
\begin{equation}
 \large 
\delta\sigma = \sigma - \sigma_c =\dot\gamma^{\frac{\omega H}{\omega H+\omega-H}} 
f\left (\frac{\dot\gamma^{\frac{(\omega+1) H}{\omega H+\omega-H}}}{T}\right)
\label{escaleo}
\end{equation}
which can be inverted, and put in the more standard form
\begin{equation}
\dot\gamma=  T^\psi G\left( {(\sigma-\sigma_c) }/T^{1/\alpha}\right)
\label{otravez7}
\end{equation}
with
\begin{equation}
\psi = \frac{\omega H+\omega-H}{(\omega+1)H}.
\label{otravez7b}
\end{equation}
\begin{equation}
\alpha = 1+\frac{1}{\omega}.
\label{otravez7c}
\end{equation}
The latter is nothing but the expression used in Eq.~(\ref{eq:yielding_thermalrounding}).

For $T\to 0$, the $T$ dependence in Eq.~\ref{otravez7} must drop out, and we get
\begin{eqnarray}
\dot\gamma= (\sigma-\sigma_c) ^\beta
\end{eqnarray}
with
\begin{equation}
\beta=\psi \alpha=1+\frac 1H-\frac 1\omega.
\end{equation}

In Figs.~\ref{fig:1pla1} and \ref{fig:1pla2} we have checked these predictions 
for the cases $\omega=1$ and $\omega=2$. 
In order to test the scaling more generally, for different values of $\omega$ we can 
use \cite{purrello2017}
\begin{equation}
 \large
-V'(x)=\frac{[1-\cos (2 \pi x)]^{\omega / 2}}{\frac{2^{\omega / 2} \Gamma\left(\omega / 2+\frac{1}{2}\right)}{\sqrt{\pi \Gamma(\omega / 2+1)}}}-1,
\label{eq:generalV}
\end{equation}
which behaves as $-V'(x)\sim A |x|^\omega$ near the transition point $x=0$, 
with $\sigma_c=1$. 
In Fig.\ref{fig:1pla1intermedia} we show that our scaling prediction for $\omega=3/2$ (Eqs.\eq{otravez7}, \eq{otravez7b} and \eq{otravez7c}), intermediate value between those corresponding to the standard $\omega=1$ and $\omega=2$ cases, is well satisfied by the data numerically generated from Eqs.\eq{app1}, \eq{app2} and \eq{eq:generalV}.

\bibliographystyle{apsrev4-2}
\bibliography{biblio}

\end{document}